\documentclass[preprint2]{proto}
\usepackage{times}

\newcommand{\refs}{\par\noindent\hangindent=1pc\hangafter=1}
\begin{document}

\title{\textbf{\LARGE Not Alone: Tracing the Origins of Very Low Mass Stars and Brown Dwarfs Through Multiplicity Studies}}

\author {\textbf{\large Adam J.\ Burgasser}}
\affil{\small\em Massachusetts Institute of Technology}
\author {\textbf{\large I.\ Neill Reid}}
\affil{\small\em Space Telescope Science Institute}
\author {{\textbf{\large Nick Siegler}} {\small and} {\textbf{\large Laird Close}}}
\affil{\small\em University of Arizona}
\author {\textbf{\large Peter Allen}}
\affil{\small\em Pennsylvania State University}
\author {\textbf{\large Patrick Lowrance}}
\affil{\small\em Spitzer Science Center}
\author {\textbf{\large John Gizis}}
\affil{\small\em University of Delaware}

\begin{abstract}
\baselineskip = 11pt
\leftskip = 0.65in
\rightskip = 0.65in
\parindent = 1pc
\small The properties of multiple stellar
systems have long provided important empirical constraints for star
formation theories, enabling (along with several other lines of
evidence) a concrete, qualitative picture of the birth and
early evolution of normal stars.
At very low masses (VLM; M $\lesssim$ 0.1~M$_{\sun}$),
down to and below the hydrogen burning minimum
mass, our understanding of formation processes is
not as clear, with several competing theories now under
consideration.  One means of testing these theories is through the
empirical characterization of VLM multiple systems.
Here, we review the results of various
VLM multiplicity studies to date.
These systems can be generally characterized as
closely separated (93\% have projected separations $\Delta <$ 20~AU) and near equal-mass
(77\% have M$_2$/M$_1$ $\ge$
0.8) occurring infrequently (perhaps 10--30\%).
Both the frequency and
maximum separation of stellar and
brown dwarf binaries steadily
decrease for lower system masses, suggesting that VLM binary
formation and/or evolution may be a mass-dependent process.
There is evidence for a fairly rapid
decline in the number of loosely-bound systems
below $\sim$0.3~M$_{\sun}$,
corresponding to a factor of 10--20 increase in the minimum binding
energy of VLM binaries as compared to more massive stellar binaries.
This wide-separation ``desert'' is present among both field ($\sim$1--5~Gyr)
and older ($>$ 100~Myr) cluster systems,
while the youngest ($\lesssim$10~Myr) VLM binaries, particularly those in nearby,
low-density star forming regions,
appear to have somewhat different systemic properties.
We compare these empirical trends
to predictions laid out by current formation theories, and outline
future observational studies needed to probe the full parameter space of the
lowest mass multiple systems. \\~\\~\\~

\end{abstract}

\section{\textbf{INTRODUCTION}}

The frequency of multiple systems and their properties
are key constraints for studies of stellar formation
and evolution. Binary and multiple stars are common in the Galaxy,
and the physical properties of the components in these systems
can be significantly influenced by dynamical and co-evolutionary processes.
Furthermore, successful theories of star formation must take into account
the creation of multiples and empirical multiplicity trends
as functions of mass, age and
metallicity.

The main focus of this review is multiplicity in very
low mass (VLM; M $\lesssim$ 0.1~M$_{\sun}$)
stars and brown dwarfs. However, to put these results
in the proper context, we start with a brief review of our
current understanding of multiplicity among higher mass stars (also see chapter
by Duch{\^{e}}ne et al.).
The standard references for binary frequency are {\it Duquennoy \& Mayor} (1991, hereafter DM91;
also {\it Abt and Levy}, 1976; {\it Abt}, 1978; {\it Mayor et al.}, 1992)
for solar-type stars and {\it Fischer and Marcy} (1992, hereafter FM92; also {\it Henry and McCarthy}, 1990;
{\it Reid and Gizis}, 1997a; {\it Halbwachs et al.}, 2003; {\it Delfosse et al.}, 2004)
for early-type M dwarfs. The DM91
survey combined spectroscopic, astrometric and direct imaging of 164 G dwarfs;
44\% of those stars were identified as binaries, with incompleteness corrections
increasing the binary fraction to $f_{bin}\sim65\%$. These corrections
include 8\% attributed to VLM companions; as
discussed further below, more recent observations show that the actual correction
is much lower. The FM92 survey covered 72 M2-M5 dwarfs within 20 parsecs, and
derived $f_{bin}=42\pm9$\%, significantly lower than the DM91 G-dwarf survey.
While both surveys include nearby stars, neither comprises a {\em volume-complete}
sample.

\begin{figure}[h]
 \plotone{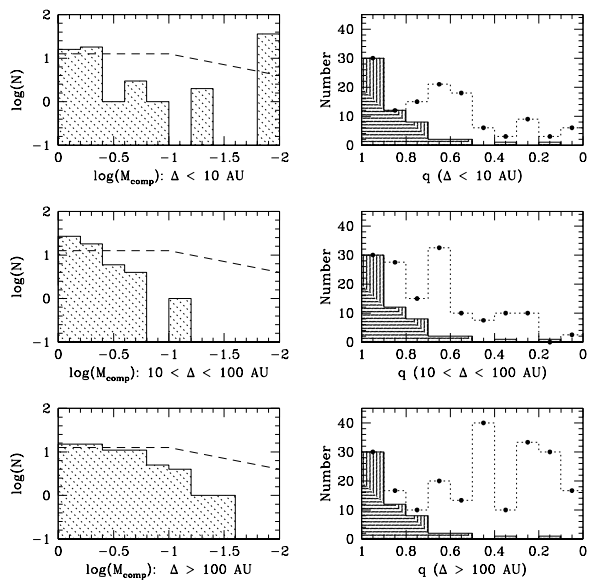}
 \caption{\small Mass and mass ratio distributions of companions to late-F to K-type dwarfs within 25
parsecs of the Sun ({\it Reid et al.}, 2002a), segregated by projected
separation/orbital semimajor axis. The left panels plot the mass
distribution of companions, with the dashed lines providing a
schematic representation of the initial mass function. The right
panels plot the mass-ratio distributions (dotted histograms), with the
solid histogram showing the mass ratio distribution
for VLM dwarfs (no segregation of separations; see Fig.~3).
These distributions are normalized at the $q = 1$ bin.
\label{fig_fk}}
 \end{figure}

Recent surveys of solar-type stars have concentrated on VLM companions.
Radial velocity (RV) surveys (e.g., {\it Marcy and Butler}, 2000; {\it Udry et al.}, 2003)
have shown that less
than 0.5\% of solar-type stars have brown dwarf companions within $\sim5$ AU. {\it Guenther et al.}
(2005) find $f_{bin}^{BD} < 2\%$ for projected
separations $\Delta<$~8~AU among Hyades stars; this is in contrast
with $f_{bin}\sim13\%$ for stellar-mass companions at those separations (DM91). At
larger separations, imaging surveys of young Solar neighborhood stars (members of the
TW Hydrae, Tucanae, Horologium and $\beta$ Pic associations; {\it Zuckerman and Song}, 2004) find
$f_{bin}^{BD} \sim6\pm4\%$ for $\Delta > 50$~AU ({\it Neuh\"auser et al.}
2003), similar to the brown dwarf companion fraction measured for field stars for separations
of 30--1600~AU ({\it Metchev}, 2005).  These fractions are $\sim$3 times lower than the hydrogen-burning
companion rate over the same separation range.
At the widest separations ($\Delta > 1000$ AU), {\it Gizis et al.}~(2001) find that solar-type
stars have comparable numbers of brown dwarf and M dwarf companions, although
this result is based on a very small number of VLM companions.

Besides the overall binary fraction, the mass distribution of companions sets
constraints on formation models. Fig.~1 shows the results for late-F to K
stars ($0.5 < (B-V) < 1.0$) within 25 parsecs of the Sun, breaking down the
sample by projected separation/orbital semi-major axis.
The left panels compare the mass distribution of companions
against a schematic representation of the initial mass function
({\it Reid et al.}, 1999, 2002a);
the right
panels compare the mass ratio ($q \equiv$ M$_2$/M$_1$)
distributions against the
VLM dwarf data assembled in this review (cf., Fig.~\ref{fig_qdist}). Clearly, low $q$
binary systems are more common at all separations among solar-
type stars than in VLM dwarfs. We return to this
issue in Section 2.2.3.
At small separations
($\Delta < 10$~AU), there is an obvious deficit of low-mass companions
(with the exception of planetary companions) as
compared to the distribution expected for random selection from the field-star
mass function.  The notorious brown dwarf desert (e.g., {\it Marcy and Butler}, 2000)
extends well into the M dwarf regime. This result is consistent with the original analysis
of {\it Mazeh and Goldberg} (1992) of the mass ratio distribution of spectroscopic
binaries, although their more recent study of proper motion
stars ({\it Goldberg et al.}, 2003) finds a bimodal distribution, with
peaks at $q\sim$0.8 and $\sim$0.2 (see also {\it Halbwachs et al.}, 2003).
The deficit in low-mass companions is less
pronounced at intermediate separations, while it
is possible that observational selection effects (e.g., sensitivity limitations)
might account for the small discrepancy for $q < 0.2$ in the wide-binary sample.

In the case of M dwarfs, attention has focused on the nearest stars. {\it Delfosse et
al.}~(2004) recently completed a spectroscopic and adaptive optics (AO) imaging
survey of M dwarfs within 9 parsecs
that is effectively complete for stellar mass companions. Combining their
results with the imaging surveys by {\it Oppenheimer et al.}~(2001) and {\it Hinz et al.}
(2002), they derive an overall binary fraction of 26\% for M dwarfs. For a more
detailed breakdown with spectral type, we can turn to the northern 8-parsec
sample
({\it Reid and Gizis}, 1997a; {\it Reid et al.}, 2003). Those data indicate
binary fractions of $24^{+13}_{-7}$\% for
spectral types M0-M2.5 (4/17 systems), $27^{+5}_{-7}$\% for M3-M4.5 (12/45 systems) and
$31^{+13}_{-9}$\% for M5-M9 (5/16 systems; uncertainties assume a binomial distribution),
where the spectral type refers to the
primary star in the system; the overall binary frequency is $f_{bin} =
27^{+5}_{-4}\%$. These results, based on volume-limited samples,
confirm that M dwarfs have significantly lower
multiplicity than more massive solar-type stars.\footnote{Even with 30\% binarity
for M dwarfs, most stars still reside in multiple systems.  As a numerical example,
consider a volume-limited sample of 100 stellar systems: 20 are type G or earlier,
10 are type K and 70 are type M.  Assuming binary fractions of 70\%, 50\% and 30\%,
respectively, these 100 systems include 140 stars, 80 in binaries and 60 in
isolated systems.  Higher order multiples only serves to increase the
companion fraction.}
This is consistent with an overall trend of decreasing multiplicity
with decreasing mass (cf., A- and B-stars have overall multiplicity
fractions as high as 80\%; {\it Shatsky and Tokovinin}, 2002; {\it Kouwenhoven et al.}, 2005).
These changes in multiplicity properties with mass among hydrogen-burning stars
emphasize that we must consider VLM dwarfs
as part of a continuum, not as a distinct species unto themselves.

\section{\textbf{OBSERVATIONS OF VERY LOW MASS BINARIES}}

\bigskip
\noindent
\textbf{2.1 Very Low Mass Binary Systems}
\bigskip

With the discovery of hundreds of VLM
dwarf stars and brown dwarfs over the past decade (see reviews by
{\it Basri}, 2000; {\it Oppenheimer et al.}, 2000; and {\it Kirkpatrick}, 2005),
it is now possible to examine systems with primaries down to
100 times less massive than the Sun. In this regime, formation mechanisms are under considerable
debate (see chapters by Bonnell et al., Goodwin et al., Klein et al., Luhman et al., and Whitworth et al.).
Hence, accurate assessment of
the multiplicity and systemic properties of VLM stars and brown dwarfs
are essential for constraining current theoretical work.

Searches for VLM binaries --- defined here as having a total system mass
M$_{tot}$ $<0.2$ M$_{\sun}$ and primary mass
M$_1$ $<0.1$ M$_{\sun}$ (cf., {\it Siegler et al.}, 2005) --- have been conducted
predominantly through high resolution imaging surveys, using both
ground-based (including
natural and, quite recently, laser guide star adaptive optics [AO]) and
space-based facilities.
Major surveys have targetted both nearby field sources
({\it Koerner et al.}, 1999; {\it Reid et al.}, 2001; {\it Bouy et al.}, 2003;
{\it Burgasser et al.}, 2003; {\it Close et al.}, 2002, 2003; {\it Gizis et al.}, 2003; {\it Siegler et al.}, 2003, 2005;
{\it Law et al.}, 2006;
{\it Allen et al.}~in preparation; {\it Bill\`eres et al.}~in preparation;
{\it Burgasser et al.}~in preparation; {\it Reid et al.}~in preparation)
and young clusters and associations ({\it Mart\'in et al.}, 1998, 2000a, 2003;
{\it Neuha\"user et al.}, 2002;
{\it Kraus et al.}, 2005; {\it Luhman et al.}, 2005; {\it Bouy et al.}, 2006).
A smaller number of high resolution spectroscopic surveys for
closely separated binaries have also taken place
({\it Basri and Mart\'in}, 1999; {\it Joergens and Guenther}, 2001; {\it Reid et al.}, 2002b;
{\it Guenther and Wuchterl}, 2003; {\it Kenyon et al.}, 2005; {\it Joergens}, 2006).
Only one eclipsing system has been discovered so far
via photometric monitoring ({\it Stassun et al.}, 2006).
Observations leading to the
identification of low mass multiple systems has been accompanied by resolved photometry
and spectroscopy, allowing characterization of the colors, luminosities
and spectral characteristics of several
binary components.
Astrometric and radial velocity monitoring has lead to mass measurements
or constraints for five VLM systems to date ({\it Basri and Mart{\'{i}}n}, 1999;
{\it Lane et al.}, 2001b; {\it Bouy et al.}, 2004a, {\it Brandner et al.}, 2004; {\it Zapatero Osorio et al.}, 2004;
{\it Stassun et al.}, 2006).

In Table 1 we list 75 VLM binary systems published in the literature
or reported to us
as of 2005. The mass criteria correspond to field dwarf
binary components later than spectral type $\sim$M6; younger systems
may include earlier spectral types.
Table~1 provides a subset of the compiled data for these sources,
given in more complete detail through an online database maintained
by N.\ Siegler (See http://paperclip.as.arizona.edu/$\sim$nsiegler/VLM\_binaries.).

\bigskip
\noindent
\textbf{2.2 General Properties of VLM Binaries}
\bigskip

Large-scale, high resolution imaging surveys in the field
have converged to similar
conclusions on the general properties of VLM field binaries.  Compared
to their higher mass stellar counterparts, VLM binaries
are
\begin{itemize}
\item rarer
($f_{bin} \approx 10-30$\%; however, see discussion below);
\item more closely separated (93\% have ${\Delta} < 20$~AU);
\item and more frequently in near-equal mass configurations (77\% have $q \ge 0.8$).
\end{itemize}
Analogous imaging surveys in young open clusters (e.g., Pleiades, $\alpha$ Persei)
find similar trends, although the youngest ($\lesssim$10~Myr)
associations (e.g., Chamaeleon~I, Upper Scorpius, Orion)
appear to exhibit somewhat different properties.  We discuss
these broad characterizations in detail below.

\bigskip
\noindent
{2.2.1 The Binary Fraction}
\bigskip

Magnitude-limited imaging surveys for VLM stars and brown dwarfs in the field
with spectral types M6 and later have
generally yielded {\it observed} binary fractions
of $\sim$20\%; taking into consideration selection effects (e.g., {\it Burgasser et al.}, 2003)
lowers this fraction to 7--15\% for $\Delta \gtrsim 2-3$~AU and
$q \gtrsim$ 0.4--0.5 ({\it Bouy et al.}, 2003;
{\it Burgasser et al} 2003; {\it Close et al.}, 2003; {\it Siegler et al.}, 2005).
{\it Burgasser et al.}~(2003) deduced $f_{bin} = 9^{+11}_{-4}$\%
for a small sample of L and T dwarfs using the $1/V_{max}$ technique ({\it Schmidt et al.}, 1968);
{\it Bouy et al.}~(2003) deduced a volume-limited
fraction of $f_{bin}$ $\sim$ 15\%.
Over the same separation ($\Delta > 2$~AU) and
mass ratio ($q > 0.5$) ranges, these multiplicity rates are less than half of those
of M dwarfs (FM92; {\it Close et al.}, 2003) and G dwarfs (DM91; {\it Bouy et al.}, 2003).
Similarly, $HST$ imaging surveys of the
125~Myr Pleiades open cluster ({\it Mart\'in et al.}, 2000a, 2003;
{\it Bouy et al.}, 2006) found a resolved binary fraction of 13--15\% for $\Delta > 7$~AU
for components at and below the hydrogen burning limit.
On the other hand, {\it Kraus et al.}~(2005) found $f_{bin}$ = 25$^{+16}_{-8}$\%
for a small sample of 0.04--0.1~M$_{\sun}$ members of Upper Scorpius
over the range $\Delta$ = 5--18~AU, somewhat higher than, but still consistent with,
other field and open cluster results.

One problem with resolved imaging surveys is their inherent selection
against tightly bound systems ($\Delta \lesssim 2-3$~AU for the field dwarfs and nearby associations,
$\Delta \lesssim 10-15$~AU for more distant star forming regions).
Here, one must generally turn to high resolution spectroscopic surveys of VLM stars,
currently few in number and with as yet limited follow-up.
{\it Reid et al.}~(2002b)
deduced a double-lined spectroscopic binary (SB2) fraction of 6$^{+7}_{-2}$\% for a sample
of M7-M9.5 field dwarfs.
{\it Guenther and Wuchterl} (2003) identified two SB2s and marginally significant RV variations in
the active M9 LP~944-20 (which they attribute to either the presence of a low-mass companion
or magnetic-induced activity) in a sample of 25 M5.5-L1.5 field and cluster dwarfs.
Including all three objects implies an observed binary fraction
of 12$^{+10}_{-4}$\%, although this value does not take into consideration selection biases.
{\it Joergens} (2006) detected one RV variable, the M6.5 Cha H$\alpha$8, among a sample
of 9 VLM stars and brown dwarfs in the 2~Myr Cha I association, implying an observed fraction
of 11$^{+18}_{-4}$\%, again subject to sampling and selection biases.
{\it Kenyon et al.}~(2005) identified four possible spectroscopic binaries (SBs) among VLM stars and brown dwarfs
in the 3-7~Myr $\sigma$ Orionis cluster on the basis of RV variations over two nights.
They derive $f_{bin} > 7-17$\% for $\Delta <$ 1~AU
(after correcting for selection effects) and a best-fit fraction of 7--19\%
(for their Sample A) depending on the assumed underlying separation distribution.
However, none of the sources from this particular study have had sufficient follow up to verify RV
variability, and cluster membership for some of the targets have been called into question.
A more thorough analysis of sensitivity and sampling biases in these SB studies
has been done by {\it Maxted and Jeffries} (2005), who find $f_{bin}$ = 17--30\%
for $\Delta < 2.6$~AU, and an
overall binary fraction of 32--45\% (assuming $f_{bin}$ = 15\% for $\Delta > 2.6$~AU).
This result suggests that imaging studies may be missing a significant
fraction of VLM systems hiding in tightly-separated pairs.  However, as orbital properties
have only been determined for two SB systems so far (PPl 15, {\it Basri and Mart\'in}, 1999; and
2MASS~0535-0546, {\it Stassun et al.}, 2006), individual separations and mass ratios for most VLM SB binaries
remain largely unconstrained.

Two recent studies ({\it Pinfield et al.}, 2003;
{\it Chapelle et al.}, 2005) have examined the fraction of unresolved (overluminous) binary
candidates among VLM stars and brown dwarfs in young associations.
Contrary to other studies,
these groups find much larger binary fractions, as high as 50\% in the {\it Pinfield et al.}
study of the Pleiades and Praesepe for $q > 0.65$.  This study also finds a binary fraction
that increases with decreasing mass, in disagreement with results in the field (see below);
the {\it Chapelle et al.}~study finds evidence for the opposite effect in the 0.9~Gyr
Praesepe cluster.  Both studies have been controversial due to the lack of
membership confirmation, and hence likelihood of
contamination;
and the possible influence of variability on the identification of overluminous sources.
Nevertheless, both of these studies and the SB results suggest that
a higher VLM binary fraction than that inferred from imaging
studies, perhaps 30\% or more, is possible.

\bigskip
\noindent
{2.2.2 The Separation Distribution}
\bigskip

\begin{figure}[h]
\includegraphics[angle=270,width=\columnwidth]{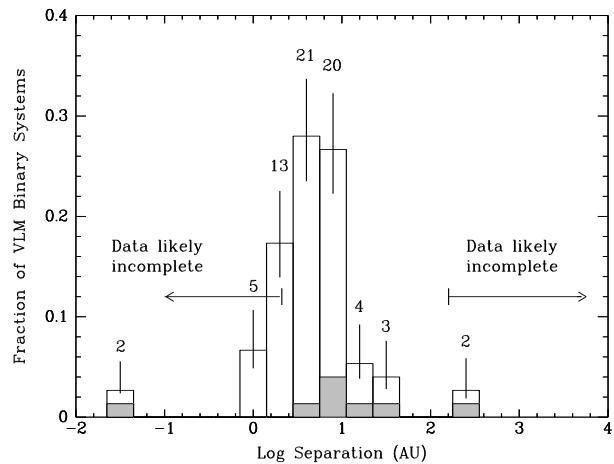}
\caption{Distribution of separations/orbital semimajor axes for known VLM binary
systems (Table 1).
The number of VLM binary systems in each 0.3~dex bin is labelled, and
uncertainties (vertical lines) are derived from a binomial distribution.
Note that SBs with unknown separations are not plotted
but included in the total number of binaries for scaling the distribution.
The distribution peaks at $\Delta \sim 3-10$\,AU, with steep declines at shorter
and longer separations.
While there is likely observational incompleteness
for $\Delta \lesssim 3$\,AU, the sharp drop in binary systems with
$\Delta \gtrsim 20$\,AU is a real, statistically robust feature.
The shaded bins represent the 8
systems with ages $<10$\,Myr. 
While the statistics are
still small, the separation distribution of these young binaries is flatter, and
suggests a peak at wider separations than that
of the field and older cluster binaries.
\label{fig_sepdist} }
\end{figure}

Fig.~\ref{fig_sepdist} plots the histogram of projected separations/orbital
semimajor
axes for 70 binaries in Table~1 (SB systems without orbital
measurements are not included).  This distribution exhibits a clear
peak around 3--10~AU, with 53$\pm$6\% of known VLM binaries encompassing
this range.
Again, because imaging surveys (from which most of the objects in Table~1 are
drawn) can only resolve systems down to a minimum angular scale (typically $0{\farcs}05-0{\farcs}1$
for AO and $HST$ programs), the
decline in this distribution at small separations is likely a
selection effect.
Results from SB
studies remain as yet unclear in this regime.
{\it Basri and Mart\'in} (1999) have suggested
that very close binaries are common based on the detection of one (PPl~15)
in a small spectral sample
The analysis of
{\it Maxted and Jeffries} (2005) suggest that there may be as many or more binaries with
$\Delta \lesssim 3$~AU as those with $\Delta \gtrsim 3$~AU.  At the extreme,
{\it Pinfield et al.}~(2003) estimate that 70--80\% of VLM binaries
in the Pleiades have $\Delta < 1$~AU, although this
result has not been corroborated by similar studies in
the Pleiades ({\it Bouy et al.}, 2006) and Praesepe ({\it Chapelle et al.}, 2005).
In any case, as the peak of the observed
separation distribution lies adjacent to the incompleteness limit,
closely separated systems
likely comprise a non-negligible
fraction of VLM binaries.

The steep decline in the separation distribution at larger separations
is, on the other hand, a statistically robust feature.
While high resolution imaging surveys are limited in this domain by field of view
(typically 10--20$\arcsec$ for $HST$ and AO studies), this only
excludes systems with $\Delta \gtrsim 150$~AU for a typical VLM field source
(distances $\sim 30$~pc) or $\Delta \gtrsim 200-1000$~AU for young cluster systems.
Even wider separations for hundreds of VLM field dwarfs should be detectable -- and are not found --
in the original surveys from which
they were identified (e.g., 2MASS, DENIS and SDSS; however, see {\it Bill\`eres et al.}, 2005).
In open clusters, deep imaging has demonstrated a consistent lack of wide companions
to VLM dwarfs.  An upper limit of $f_{bin} < 8$\% for
$\Delta > 11$~AU is derived for the 90~Myr $\alpha$ Per open cluster ({\it Mart\'in et al.}, 2003),
similar to the 5\% upper limit for $\Delta > 15$~AU measured for
32 VLM members of the 2~Myr IC~348 cluster ({\it Luhman et al.}, 2005).
{\it Lucas et al.}~(2005)
measure an upper limit of 2\% for wide VLM binaries ($\Delta > 150$~AU) in the
1~Myr Trapezium
cluster based on a two-point correlation function.
In contrast, 93\% of the known VLM binaries have $\Delta \lesssim$ 20~AU.
Hence, a ``wide brown dwarf binary desert'' is evidenced for
VLM stars and brown dwarfs ({\it Mart\'in et al.}, 2000a), a potential clue to their formation.

While survey results have generally been negative for wide VLM binaries, two
--- 2MASS~J11011926-7732383AB ({\it Luhman}, 2004; hereafter 2MASS~1101-7732AB) and
DENIS~J055146.0-443412.2AB ({\it Bill\`eres et al.}, 2005, hereafter DENIS~0551-4434AB) --
have been identified serendipitously.
These systems
have projected separations $\gtrsim$~200~AU,
over 10 times wider than the vast majority of VLM binary systems.
A third low mass binary not included in Table~1,
GG~Tau~BaBb (a.k.a.\ GG Tau/c; {\it Leinert et al.}, 1991; {\it White et al.}, 1999),
with estimated primary and total system masses of 0.12 and 0.16 M$_{\sun}$, respectively,
also has a projected separation greater than 200~AU.
Interestingly, two of these three systems are members of very young, loose associations.
We discuss these source further in $\S$2.4.2.

The separation distribution of VLM stars therefore peaks at or below $\sim$3--10~AU,
corresponding to orbital periods of $\lesssim$40~yr.  This is quite different from the
separation distribution of G dwarfs, which shows a broad peak around 30~AU (periods of $\sim$170~yr; DM91);
and the M dwarf distribution, which peaks between 4--30~AU (periods of 9--270~yr; FM92).
There is a suggestion in this trend of decreasing separations as a function of mass,
as discussed further below.

\bigskip
\noindent
{2.2.3 The Mass Ratio Distribution}
\bigskip

\begin{figure}[h]
\includegraphics[angle=270,width=\columnwidth]{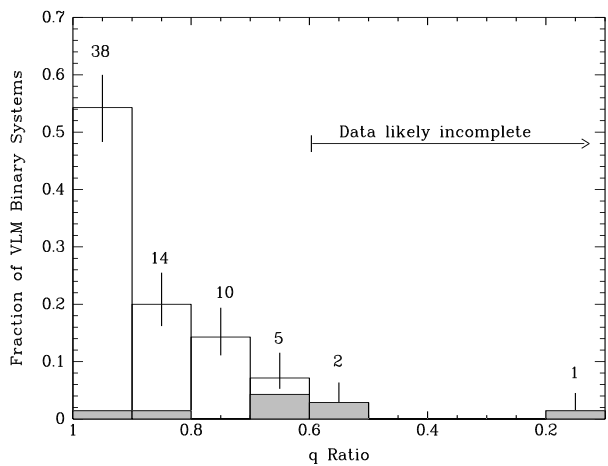}
\caption{Mass ratio distribution of known VLM binary
systems (Table 1).
The number of VLM binary systems in each 0.1 fractional
bin is labelled, and uncertainties are derived from a binomial distribution.
Note that SBs with unknown mass ratios are not plotted
and not included in the total number of binaries when scaling the distribution.
The distribution peaks near unity for binary
systems with
$\Delta \gtrsim 3-4\,$AU, and matches a power law.
Note that incompleteness is likely for $q \lesssim 0.6$.
The shaded bins
represent the eight systems with ages $<10$\,Myr.
While the statistics are still small, the mass ratio distribution of these
young systems suggests a flatter distribution
than that of field and older cluster binaries.
\label{fig_qdist}}
\end{figure}

Fig.~\ref{fig_qdist} shows the distribution of mass ratios for
70 of the binaries in Table~1 (not including
SBs without mass estimates).
These ratios were derived by a variety of methods, including
comparison of component fluxes to evolutionary models (e.g., {\it Chabrier et al.}, 2000),
analytic relations (e.g., {\it Burrows et al.}, 2001)
and direct estimates from orbital motion measurements.  Despite these
different techniques, a comparison of all the data shows congruence
with individual studies.  The mass ratio distribution for VLM systems
is strongly peaked at near-unity ratios; over half of the known VLM binaries
have $q > 0.9$ and 77$^{+4}_{-5}$\% have $q \ge 0.8$.

As with the separation distribution, it is important to consider selection effects
in the observed mass ratios.
Most pertinent is the detectability of secondaries in
low $q$ binaries, which may be too faint
for direct imaging or of insufficient mass to induce a measureable RV
variation in the primary's spectrum.  The former case is an important issue
for field binaries, as low mass substellar companions fade to obscurity over time.
However, most imaging and spectroscopic surveys to date are sensitive down to $q \gtrsim 0.5$,
while a sharp dropoff is clearly evident at the highest mass ratios.
Hence, while the number of low mass ratio systems may be underestimated,
the $q \sim 1$ peak is not the result of this bias.

A second effect is the preferential discovery of unresolved
equal-mass systems in wide field surveys.
As such systems are twice as bright as their single counterparts, they are $\sim$3 times
more likely to be found than single sources in a magnitude-limited survey.
Systems with lower mass ratios are not as overluminous and less affected by this bias.
{\it Burgasser et al.}~(2003) examined this
impact of this bias on a small sample of L and T dwarf binaries and found it to be
significant only for $q \lesssim 0.6$.  Hence, this bias cannot be responsible for
the $q \sim 1$ peak.

VLM (field and open cluster) binaries therefore show a clear
preference for equal mass systems, in contrast to the majority of
F--K stellar systems (Fig.~\ref{fig_fk}).
It is worth noting that M dwarfs in the 8~pc sample show a similar, although less pronounced, $q \sim 1$ peak
({\it Reid and Gizis}, 1997a), again suggesting a mass-dependent trend.

\bigskip
\noindent
{2.2.4 Higher Order Multiples}
\bigskip

Thus far we have focused on VLM binaries, but higher order multiples (triples,
quadruples, etc.) are also abundant among more massive
stars, comprising perhaps 15--25\% of all multiple stellar systems ({\it Tokovinin}, 2004;
see chapter by Duch\^ene et al.).  Several VLM binaries are
components of higher order multiple systems with
more massive stars.
{\it Burgasser et al.}~(2005a) have even suggested a
higher binary fraction for brown dwarfs that are widely-separated
companions to massive stars.
Higher order multiples are currently rare among purely VLM systems, however.
The LP 213-67/LP 213-68AB system is one exception, with the three components
(spectral types M6.5, M8 and L0) forming a wide hierarchical triple with separations
of 340~AU and 2.8~AU ({\it Gizis et al.}, 2000a; {\it Close et al.}, 2003).
DENIS~0205-1159AB may also have a third component, marginally resolved through
high resolution imaging ({\it Bouy et al.}, 2005).  Considering both systems
as VLM triples, the ratio of high-order multiples to binaries is only 3$^{+4}_{-1}$\%,
quite low in comparison to higher mass stars.  This may be due to selection effects,
however, as the already tight separations of VLM binaries implies that the
third component of a (stable) hierarchical triple
must be squeezed into an extremely small orbit.  Indeed, this could
argue against a large fraction of higher order VLM systems.  On the other hand,
undiscovered wide tertiaries (as in LP 213-67/LP 213-68AB)
may be present around some of these systems.  Additional
observational work
is needed to determine whether higher order VLM multiples are truly less common
than their stellar counterparts.

\bigskip
\noindent
\textbf{2.3 Statistical Analysis: Bayesian Modeling}
\bigskip

To examine the observed binary properties of resolved VLM stars in more detail,
we performed a Bayesian statistical analysis of
imaging surveys to date.
The Bayesian approach allows the
incorporation of many disparate data sets, and the
easy assimilation of non-detections,
into a unified analysis of a
single problem ({\it Sivia}, 1996).
We focused our analysis on the surveys of
{\it Koerner et al.}~(1999); {\it Reid et al.}~(2001); {\it Bouy et al.}~(2003); {\it Close
et al.}~(2003); {\it Gizis et al.}~(2003); {\it Siegler et al.}~(2005); and
{\it Allen et al.}~(in preparation).
The Bayesian statistical method employed is similar to that described
in {\it Allen et al.}~(2005).

We first
constructed a set
of parameterized companion distribution models in terms of orbital
semi-major axis ($a$) and companion mass ratio.  For the
semi-major axis distribution we use a Gaussian in log AU given by:
\begin{equation}
P(a_{0},\sigma_a)=\frac{1}{\sqrt{2{\sigma_a}^{2}}}e^{-({\log}(a)-{\log}(a_0))^{2}/2{\sigma_a^2}}
\end{equation}
where $a_0$ is the peak of the Gaussian and $\sigma_a$
is the logarithmic half-width, both variable parameters.
This formulation is prompted by the results of DM91 and FM92
(however, see {\it Maxted and Jeffries}, 2005).
For the mass-ratio model, we assume
a power law of
the form:
\begin{equation}
P(N,\gamma) = N\frac{q^{\gamma}}{{\int_0^1}q^{\gamma}}
\end{equation}
where the normalization factor $N$ is defined to be
the overall binary fraction (i.e., $f_{bin}$), and $\gamma$ is a variable parameter.

In order to compare the model distributions to the data, we transform them to
observables, namely the log of the projected separation
($\log{\Delta}$) and the difference in magnitude between the
secondary and the primary (${\Delta}M$).
The former is computed by transforming the
semi-major axis distribution as:
\begin{equation}
\Delta = a\sqrt{cos^{2}(\phi)sin^{2}(i) + sin^{2}(\phi)},
\end{equation}
where we assume a uniform distribution of
circular orbits over all possible inclinations $(i)$ and phases
($\phi$).
The transformation of the $q$ distribution to a ${\Delta}M$ distribution
is done by
assigning each mass ratio a range of possible luminosities for ages between 10 Myr and 10 Gyr
using evolutionary models from {\it Burrows et al.}~(2001).

The transformed model distributions are then compared to the
observed distributions via a
Bayesian statistical method, as described in
{\it Allen et al.}~(2005).
The models are directly compared to the data after being
convolved with a window function, which describes how many times a bin
in observational space ($\Delta$, ${\Delta}M$) was observed in a
particular survey.  In this way we do not analyze the models
where there is no data, and the relative frequency of observations
is taken into account.

\begin{figure}[ht]
\begin{center}
\includegraphics[scale=0.65,angle=270]{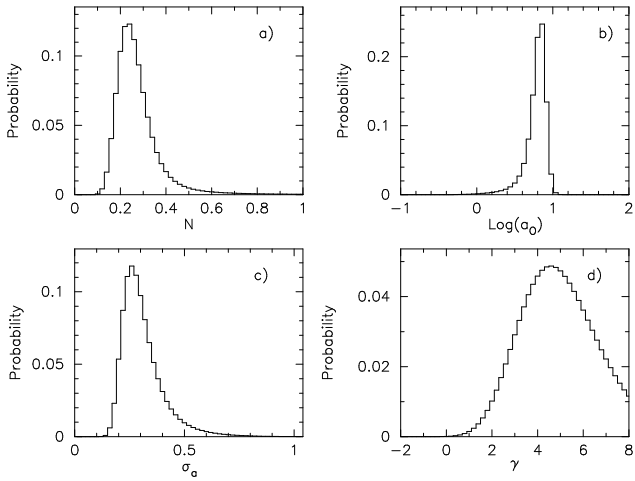}
\end{center}
\caption{Posterior probability distributions of the four companion
model parameters: a) overall binary fraction (N); b) center of
the semi-major axis distribution ($\log(a_0)$); c) width of the
semi-major axis distribution ($\sigma_a$); d) mass ratio distribution
power law index ($\gamma$).} \label{fig:jo}
\end{figure}

The output posterior distribution is four dimensional
($\log(a_0)$, $\sigma_a$, $N$, $\gamma$) and
is impossible to display in its entirety.  Instead, we show
marginalized distributions (Fig.~\ref{fig:jo}), collapsing the posterior distribution along
different parameter axes.
These distributions have a non-negligible dispersion,
as parameters spaces outside
the observational window function (e.g., very tight binaries) add
considerable uncertainty to the statistical model.
Nevertheless, the distributions are well-behaved and enable
us to derive best-fit values and uncertainties for the various
parameters.  The overall binary fraction is reasonably well constrained,
$N = 22^{+8}_{-4}$\%, with a long tail in its probability
distribution to higher rates. The remaining parameters are
$\log(a_0) = 0.86^{+0.06}_{-0.18}~\log(AU)$,
$\sigma_a = 0.24^{+0.08}_{-0.06}~\log(AU)$, and
$\gamma = 4.8^{+1.4}_{-1.6}$ (all listed uncertainties are 68\%
confidence level).

\begin{figure}[ht]
\begin{center}
\includegraphics[width=0.65\columnwidth,angle=270]{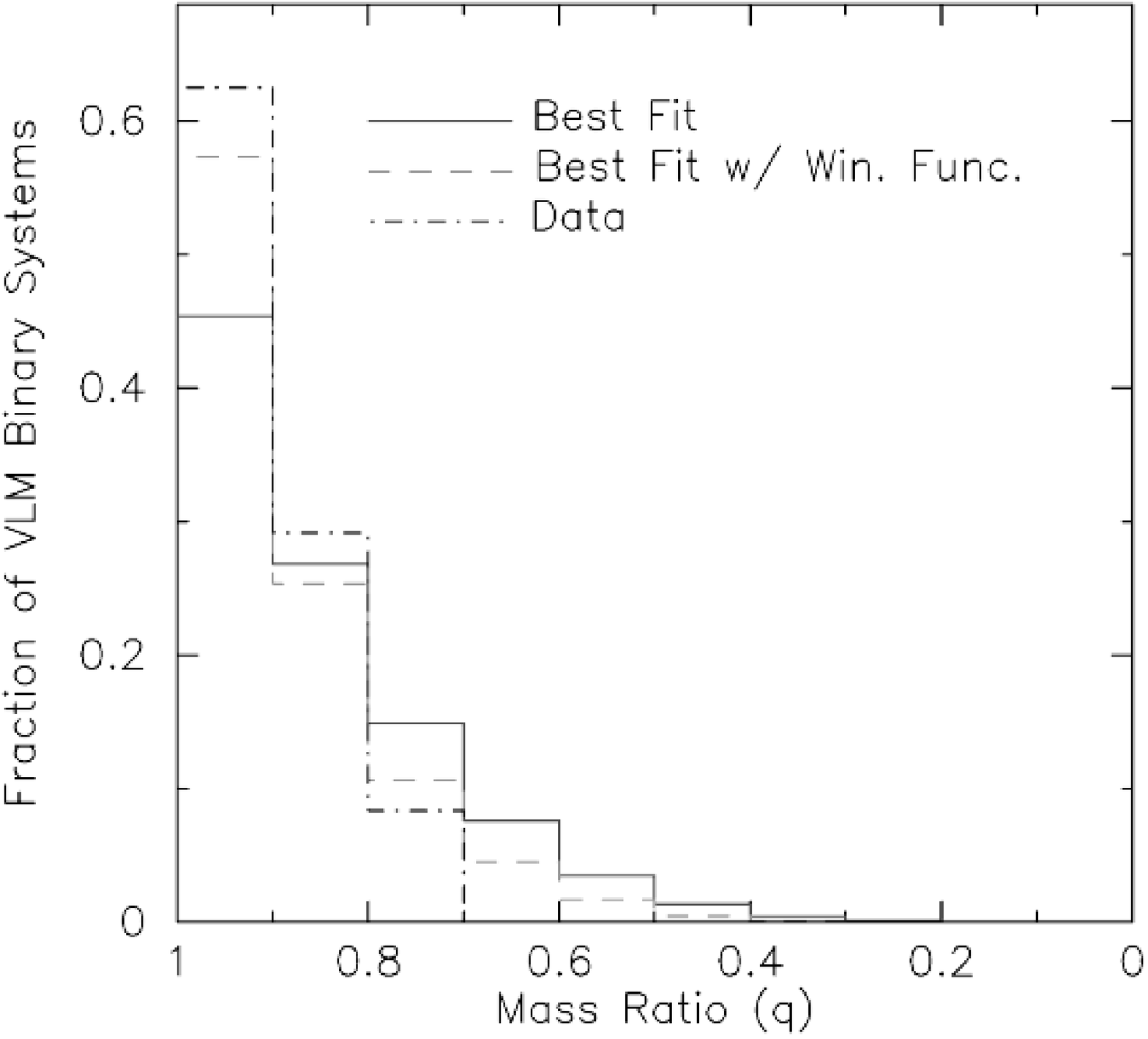}
\includegraphics[width=0.65\columnwidth,angle=270]{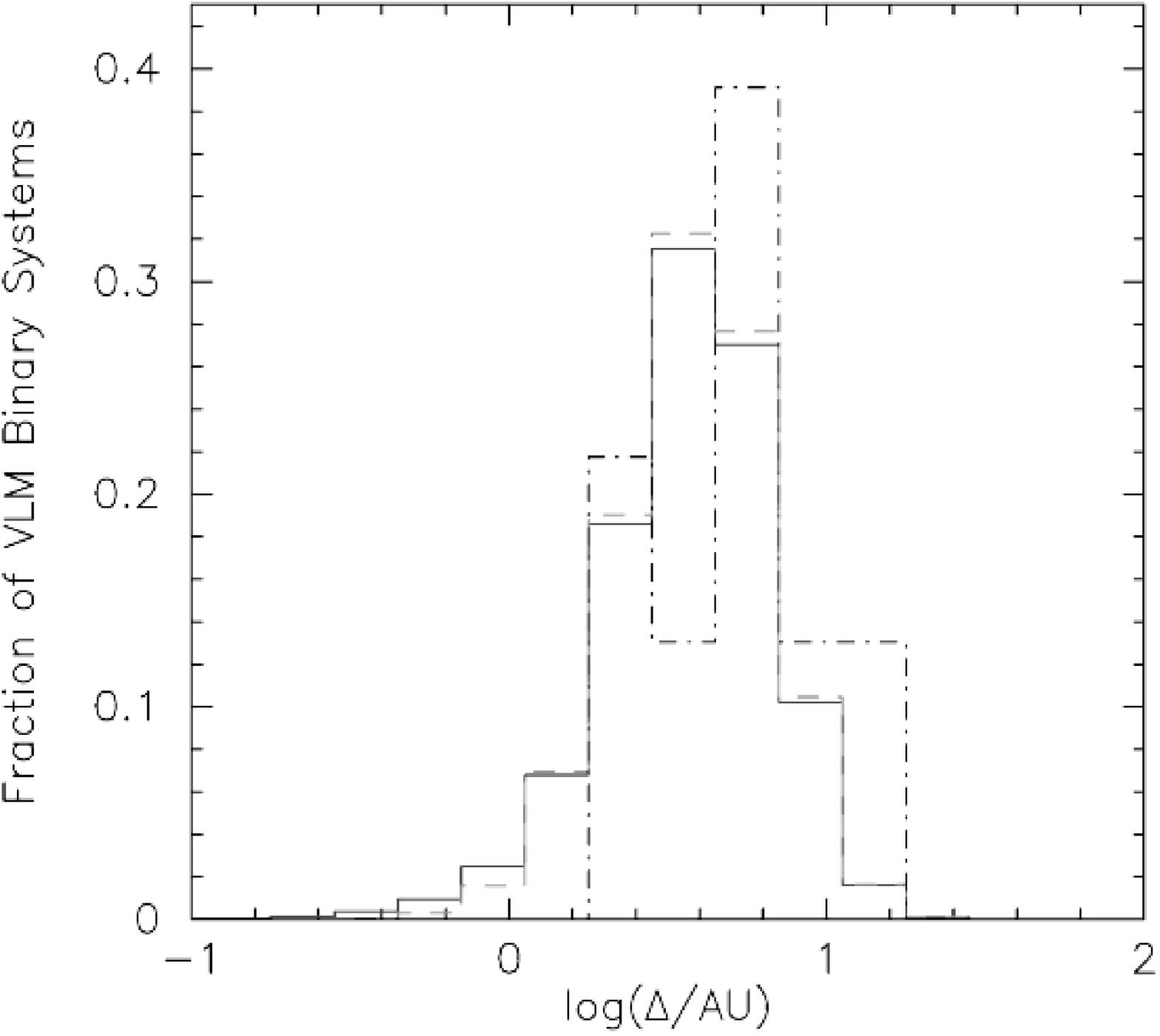}
\end{center}
\caption{ {\em (Top)} The fraction of VLM binaries with a given $q$ for the
best fit model (solid line), the best fit model view through the
window function (dashed line), and the data used in the Bayesian
analysis (dot-dashed line).  Note how the window function
over-emphasizes the high mass ratio systems.  {\em (Bottom)} The
projected separation distribution for the best fit model
(lines are the same as the top panel).
\label{fig:inom}}
\end{figure}

The mass ratio and projected separation distributions inferred from the
best-fit parameters are shown in Fig.~\ref{fig:inom}.
The best-fit binary
fraction is 22\%, but after applying our window function the expected
resolved fraction is
$\sim$17\%, slightly higher than but
consistent with the observed $f_{bin}$ from imaging surveys
($\S$~2.2.1).
The best-fit mass ratio distribution is highly peaked near $q$ = 1,
similar to the data but somewhat flatter than observed due to
selection effects in the empirical samples. This nevertheless confirms
that the mass ratio
distribution is fundamentally peaked towards high $q$ values.

The best fit value for the peak of the semimajor axis distribution is
$\sim$7~AU, implying a peak in the projected separation distribution of about
3.5~AU, matching well with the data (Fig.~\ref{fig:inom}b).  The best fit width of this distribution is quite
narrow, implying very few wide
systems ($>20~AU \sim 1\%$) and very few close systems ($<1$~AU
$\sim$ 2-3\%). It is important to stress that the imaging
data provide weak constraints on closely-separated binaries,
and the latter fraction may be somewhat higher (cf., {\it Maxted and Jeffries}, 2005).
On the other hand,
the constraint on the wide binary fraction
(1\% or less) is the most robust result of this analysis. Between
all of the surveys considered here there are over 250 unique
fields that have been probed for companions out to hundreds of AU
with no detections.
Hence, such pairings are exceptionally rare.

\bigskip
\noindent
\textbf{2.4 Discussion}
\bigskip

\bigskip
\noindent
2.4.1 On the Preference of Tight Binaries
\bigskip

\begin{figure*}[ht]
\includegraphics[angle=0,width=1.0\columnwidth]{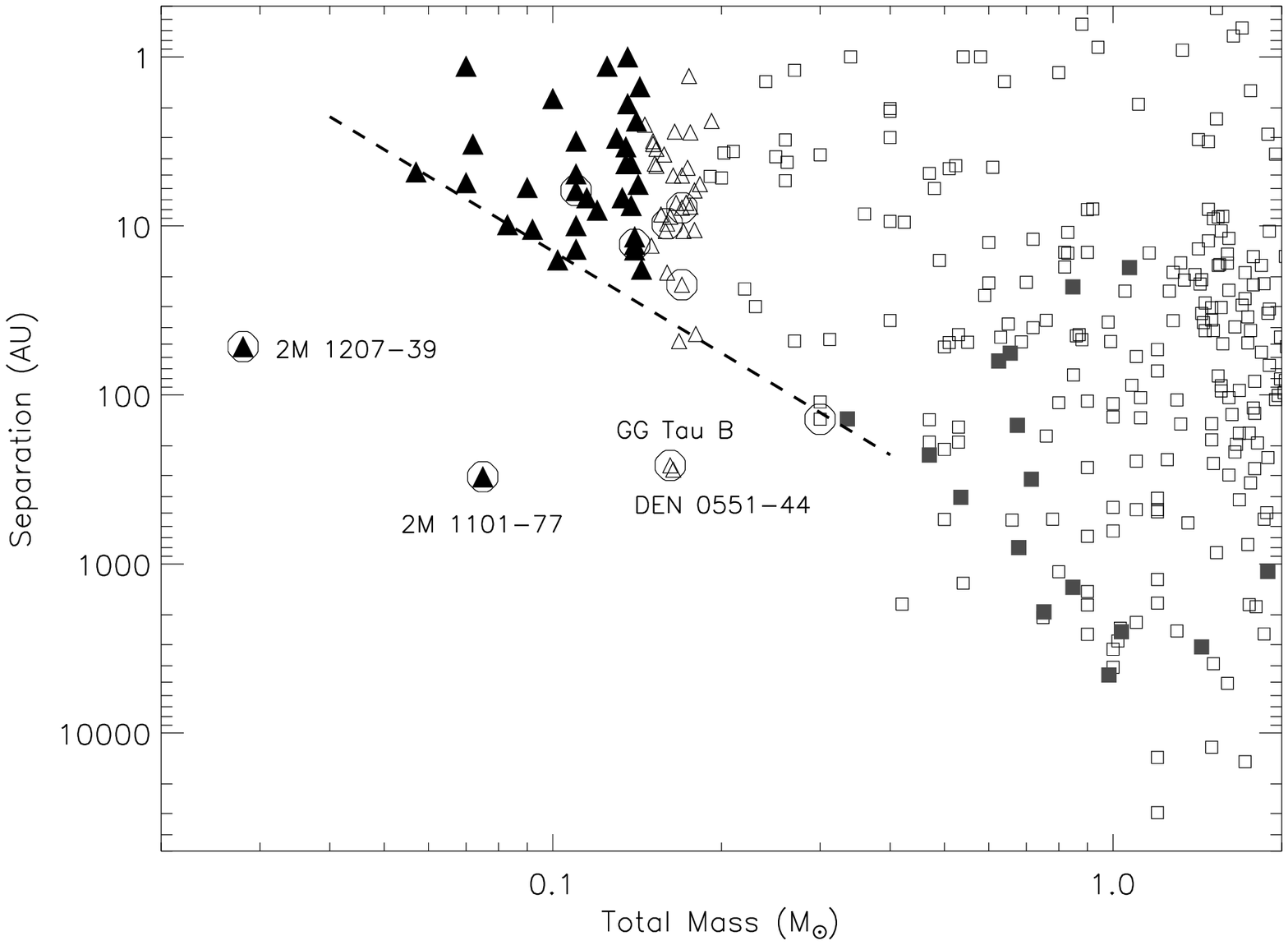}
 \includegraphics[angle=0,width=1.0\columnwidth]{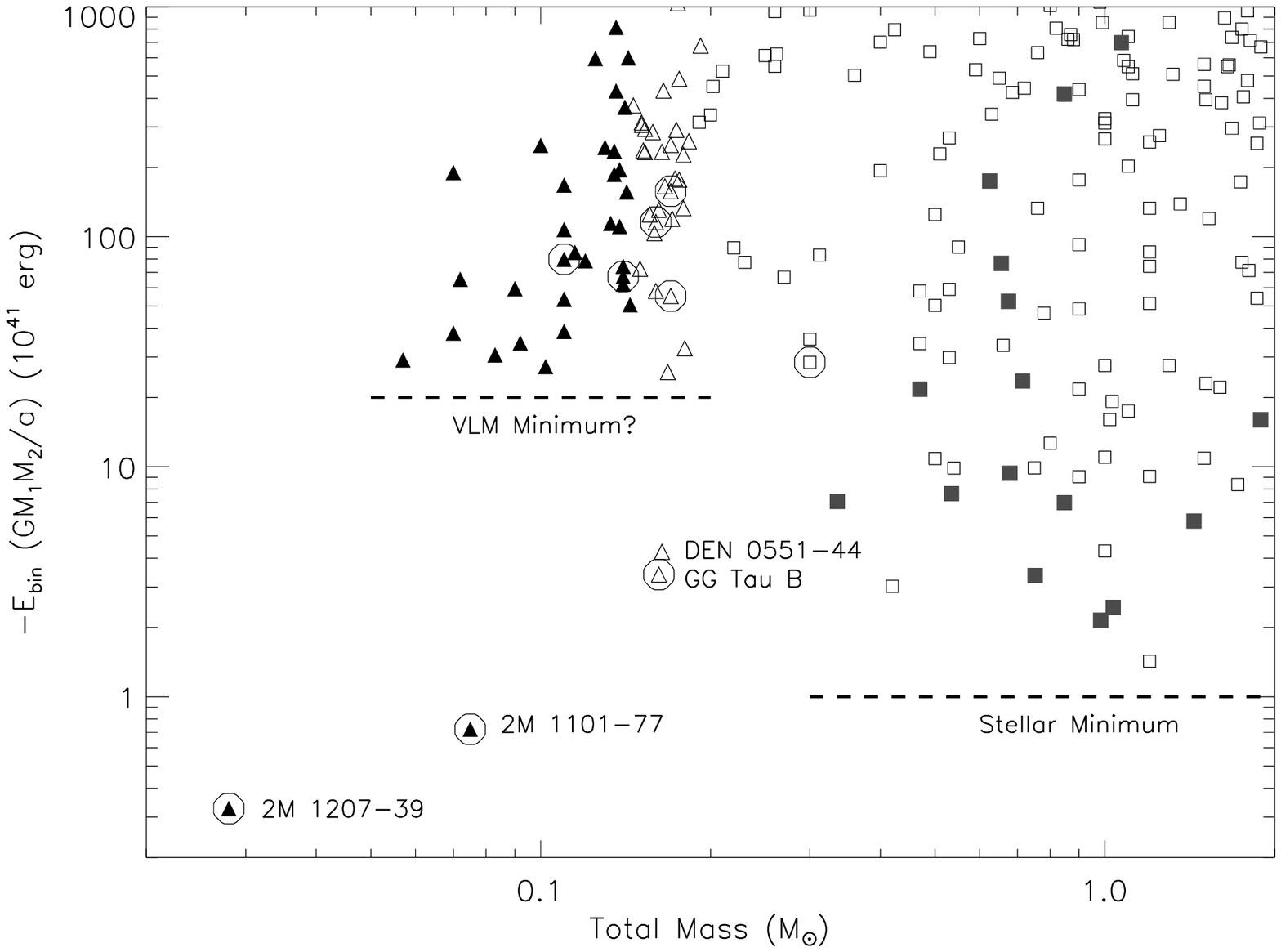}
\caption{{\em (Left)}
Separation (in AU) versus total system mass (in Solar masses) for known binary systems.
Stellar binaries from {\it Close et al.}~(1990);
DM91; FM92; {\it Reid and Gizis}, (1997b); and {\it Tokovinin}, (1997) are shown as open squares;
stellar-brown dwarf systems compiled by {\it Reid et al.}~(2001) are shown as filled squares.
The 68 binaries from Table 1 with measured projected separations and estimated masses
are plotted as triangles; filled triangles indicate substellar primaries.
Systems younger than 10 Myr are encircled.
The dotted line indicates the maximum separation/system mass relation
for VLM stellar and substellar binaries proposed by {\it Burgasser et al.}~(2003),
indicating that lower mass systems are more tightly bound (see also {\it Close et al.}, 2003).
However, three
young systems (GG~Tau~BaBb, 2MASS~1101-7732AB and 2MASS~1207-3932AB),
and the field binary DENIS~0551-4434AB, all appear to contradict these trends.
{\em (Right)} Same systems but this time comparing binding energy ($-E_{bind} = G{\rm M}_1{\rm M}_2/a$)
 to total system mass.  As first pointed out in {\it Close et al.}~(2003), the widest
 VLM field binaries are 10--20 times more tightly bound
 than the widest stellar binaries, with the singular exception of DENIS~0551-4434AB.
 On the other hand, the three young VLM
 systems GG Tau BaBb, 2MASS 1101-7732AB and
 2MASS 1207-3932AB are much more weakly bound.
\label{fig_avsmass}}
\end{figure*}

The sharp decline in the VLM binary fraction for $\Delta > 20$~AU is not a
feature shared with more massive stellar systems, which can extend from 0.1~AU
to 0.1~pc.  However, the decline is consistent with the observed
trend of smaller mean separations,
and smaller maximum separations, from A to M field binaries.
This is demonstrated in Fig.~\ref{fig_avsmass}, which plots
projected separations/semimajor axes versus total system mass for
stellar and substellar field and cluster binaries.
The maximum separations ($\Delta_{max}$) of these systems show a striking dependence on total
system mass.  Prior to the discovery of the wide pairs 2MASS~1101-7732AB and DENIS~0551-4434AB,
{\it Burgasser et al} (2003) found a power-law relation between $\Delta_{max}$ and total system mass,
$\Delta_{max}$ = 1400(M$_{tot}$/M$_{\sun})^2$~AU, that appeared to fit all
VLM systems known at that time.  Similarly, {\it Close et al.}~(2003) found a linear relation of
$\Delta_{max}$ = 23.2(M$_{tot}$/0.185~M$_{\sun}$)~AU for VLM binaries,
corresponding to a minimum escape velocity $V_{esc}$ = 3.8 km~s$^{-1}$.  This
was greater than
a minimum value of $V_{esc}$ = 0.6 km~s$^{-1}$ inferred for more massive stellar systems,
and both results indicate that lower mass binaries are progressively more tightly
bound.
{\it Close et al.}~(2003) further pointed out a possible ``break'' in the minimum
binding energies of stellar and VLM binaries, also shown in
Fig.~\ref{fig_avsmass}.  Around M$_{tot}$ $\approx$ 0.3~M$_{\sun}$, the majority of wide VLM systems
appear to be 10--20 times more strongly bound than the widest stellar systems.

More recently, exceptions to the empirical
trends shown in Fig.~\ref{fig_avsmass}  have been identified,
including the widely-separated
VLM systems 2MASS~1101-7732AB, DENIS~0551-4434AB and GG Tau BaBb.
In addition, the extremely low mass (M$_{tot}$ $\approx$ 0.03~M$_{\sun}$)
brown dwarf pair 2MASS~J12073346-3932549AB
({\it Chauvin et al.}, 2004, 2005a; hereafter 2MASS~1207-3932AB), identified in the 8~Myr TW~Hydrae moving group
({\it Gizis}, 2002), falls well outside the mass/$\Delta_{max}$ limits outlined above.
Such ``exceptions'' have called into question whether current
empirical separation limits are representative of VLM systems in general, and can be considered
robust constraints for
formation models; or if wide binaries are a normal (if rare) mode of VLM binary formation.
These questions remain under debate.

\bigskip
\noindent
2.4.2 Do Evolution or Environment Play a Role in VLM Binary Properties?
\bigskip

That three of the four weakly bound VLM systems are in young ($\lesssim$10~Myr),
low density associations may be an important clue to their formation and existence,
and encourages a closer examination of the multiplicity
properties of such objects in general.  The shaded histograms in
Figs.~\ref{fig_sepdist} and~\ref{fig_qdist}
delineate the separation and mass ratio distributions, respectively, of the 8
binaries in Table~1 that are members of clusters or associations
younger than 10~Myr.  These distributions, although based on small number statistics,
are nevertheless compelling. Young systems show a much broader range of separations,
spanning 0.04 $\lesssim \Delta \lesssim$ 240~AU, with 25$^{+19}_{-9}$\% (2/8)
having $\Delta >$ 20~AU (as compared to 5$^{+4}_{-1}$\% of older VLM systems).
The mass ratio distribution is also quite flat, with a {\it statistically
significant} shortfall in the relative number of $q \ge 0.8$ binaries (25$^{+19}_{-9}$\%
versus 81$^{+4}_{-6}$\%).  Assuming that the older field sources predominately
originate from young clusters ({\it Lada and Lada}, 2003), these differences suggest
an {\em evolution} of VLM binary properties over a timescale
of 5--10~Myr.

However, care must be taken when interpreting these data, as selection effects can
distort the underlying distributions.
Because the youngest brown dwarfs are still quite warm and luminous, imaging surveys
in young clusters can generally probe much smaller masses --- and hence smaller mass ratios --- than
equivalent surveys of older clusters or in the field.  In addition, with the exception of some
nearby moving groups (e.g., TW~Hydra, Ursa Major), most of the youngest clusters lie
at larger distances, so closely separated systems ($\Delta \lesssim 10$~AU)
cannot be generally resolved through direct imaging.
This biases young samples against the close separations typical
of field binaries.
So in fact there may be many more closely-separated young VLM pairs,
or many more widely-separated, small mass ratio older VLM pairs, than currently known.

What about older VLM members of the
Galactic thick disk and halo?  Unfortunately, current imaging searches for companions
to low-mass subdwarfs are not yet capable of detecting substellar companions directly,
and radial velocity surveys of the necessary frequency are not yet
complete.
{\it Gizis and Reid} (2000) imaged nine VLM metal-poor (M subdwarf)
primaries with HST, and found that none had companions down to the
hydrogen-burning limit.  This sample has been extended to a total of 28
M subdwarfs within 60 parsecs, but all appear single ({\it Riaz and Gizis}, in
preparation).  Taken at face value, this result ($f_{bin} < 6$\%)
suggests that halo VLM doubles
with separations in the range 5-100 AU are rarer than those in the disk
population.  However, given the danger of unknown selection biases, the possibility
of metallicity effects, and the still
small numbers of the empirical sample, this result should be taken with caution.

The current data also support the possibility
that environment may play a role in the multiple properties of VLM systems.
The three young, widely-separated binaries discussed above all reside in loose associations
that have average stellar densities
of 0.01--1~pc$^{-3}$ (e.g., {\it Luhman}, 2004; {\it Mamajek}, 2005), too low
for stellar encounters to have a significant disruptive effect
({\it Weinberg et al.}, 1987).
This is in contrast to high-density star formation regions
such as Orion, where average
densities of 10$^4$~pc$^{-3}$ ({\it Hillenbrand}, 1997) are sufficient for
stellar encounters to disrupt $\sim$10~AU VLM binaries over a $\sim$10~Myr timescale
({\it Weinberg et al.}, 1987; {\it Burgasser et al.}, 2003).
The influence of stellar density has been
cited for observed differences in multiplicity among solar-mass stars
in various clusters (e.g.,
{\it Ghez et al.}, 1993; {\it Scally et al.}, 1999; {\it Patience and Duch{\^{e}}ne}, 2001;
{\it Lada and Lada}, 2003; also see chapter by Duch{\^{e}}ne et al.), so differences
among VLM binaries should not be surprising.
This scenario can also explain the paucity of wide binaries in the field.
Dense embedded clusters, in which wide binaries can be easily disrupted
(cf., {\it Kroupa}, 1995a,b,c)
contribute perhaps 70--80\% of the stars in the Galaxy ({\it Lada and Lada}, 2003).
The few wide systems created in less dense clusters or associations would therefore
comprise a negligible fraction of all VLM binaries in the field (cf., {\it Kroupa and Bouvier}, 2003).
This scenario is compelling, but requires better statistics
to be tested sufficiently.

\section{\textbf{CONFRONTING THE MODELS}}

With a full analysis of the empirical properties of VLM
multiple systems in hand, we now examine how the predictions of
current star and brown dwarf formation theories compare.
Detailed discussion on the current modelling efforts are provided
in the chapters of Ballesteros-Paredes et al., Bate et al., Goodwin et al.,
Klein et al.\ and Whitworth et al.  Comparison of formation theories
with the general properties (mass function, disk fraction, etc.)
of low mass stars and brown dwarfs
are provided in the chapters of Duch\^ene et al.\ and Luhman et al.
Here we focus primarily
on the predictions for VLM multiplicity.

\bigskip
\noindent
\textbf{3.1 Fragmentation and Dynamical Evolution}
\bigskip

Undoubtedly, gravitational contraction of dense cores in molecular clouds
provides the fundamental building blocks for stellar and substellar objects.
However, the details of the contraction and subsequent evolution
of the cores remain critical details under considerable debate.
This is particularly the case for VLM star and brown dwarfs
whose masses are significantly below the Jeans mass ($\sim$1~M$_{\sun}$),
and as such cannot be formed efficiently through basic contraction
scenarios (e.g., {\it Shu et al.}, 1987).
The inclusion of additional physics, such as
magnetic field effects ({\it Boss}, 2001, 2002, 2004)
and turbulent fragmentation, has brought some resolution
to this problem, and has enabled a new generation of
VLM formation models.

Turbulent fragmentation
({\it Henriksen}, 1986, 1991; {\it Larson}, 1992; {\it Elmegreen}, 1997, 1999, 2000),
in which gas flows collide, are compressed
and form gravitationally unstable clumps,
has pushed the fragmentation mass limit down to the
effective opacity limit, of order 0.01 M$_{\sun}$ ({\it Bate et al.}, 2002).
{\it Boyd and Whitworth}, (2005) have
modelled the turbulent
fragmentation of two dimensional sheets and found a protostellar
mass distribution that extends
to ${\sim}0.003~M_{\odot}$.
{\it Padoan and Nordlund} (2004) and {\it Padoan et al.}~(2005) have studied
three dimensional turbulent fragmentation of a molecular cloud using an
Adaptive Mesh Refinement code, and are also
capable of
producing cores as small as
${\sim}0.003~M_{\odot}$.
In these studies, no predictions are made on the overall multiplicity of the
protostars.  However,
fragmented cores naturally lead to the creation of
gravitationally-bound, high-order multiple systems,
as confirmed in multiplicity studies of Class 0 and I
protostars ({\it Haisch et al.}, 2002, 2004; {\it Reipurth et al.}, 2002, 2004;
{\it Duch\^ene et al.}, 2004; see chapter by Duch\^ene et al.),
and therefore provide a natural framework for the creation
of VLM multiple systems.

However, it is well known that N-body groups
are generally dynamically unstable, and dynamical
scattering will dissolve such systems in a few crossing times
($\sim$10$^5$ yr), preferentially
removing the lowest-mass members (e.g., {\it Kroupa et al.}, 1999).
The scattering of low-mass bodies will also limit the
accretion of gas and dust onto initially substellar cores,
which would otherwise build up to stellar masses.
These ideas have led to the so-called ``ejection'' model
for brown dwarf formation ({\it Reipurth and Clarke}, 2001), in which brown
dwarfs (and presumably VLM stars) are simply stellar embryos
ejected from their nascent cloud.
This model has received a great deal of attention recently,
as its qualitative multiplicity predictions -- a small fraction of multiples
and a preference for strongly bound binaries (close separations and near-unity mass
ratios) -- appear to fall in line with observational results.

The most comprehensive simulations of this scenario, incorporating
both Smoothed Particle Hydrodynamics (SPH) modelling for fragmentation and
accretion and N-body simulations for dynamic interactions, have been
produced by M.\ Bate
and collaborators ({\it Bate et al.}, 2002, 2003; {\it Bate and Bonnell}, 2005), and are described
in detail in the chapter by Bate et al.
Their original simulation of a 50 M$_{\sun}$ cloud
produced only one brown
dwarf-brown dwarf binary system, still accreting and dynamically
unstable at the end of the simulation, implying a VLM binary fraction of ${\lesssim}5\%$.
It was immediately recognized that this fraction may be too low
when compared to observations ({\it Close et al.}, 2003).
Later simulations ({\it Bate and Bonnell}, 2005) found that
higher VLM binary fractions (up to 8\%) were possible in denser clouds.
The highest density simulation also
produced stable wide ($> 60$~AU)
VLM binary systems
when low-mass cores were ejected in the same direction and became
bound.  It is important to note that the two wide young VLM systems currently
known are, on the contrary, associated with {\em low-density} associations.

While the {\it Bate et al.}~simulations have provided a great leap forward
in the modelling of low mass star formation, their relevance to the
observed properties of VLM binaries are hindered by necessary computational
approximations.  First, sink particles
encompassing all bound gas within 5~AU
are used when densities exceed 10$^{-11}$ g~cm$^{-3}$.
This approximation rules out any binaries more closely separated than this limit,
encompassing a majority of VLM systems (see $\S$2.2.2).
Second, a softened Newtonian potential is employed below separations
of 5~AU (down to 1~AU), which enhances the disruption of binary pairs
with smaller separations ({\it Delgado-Donate et al.}, 2004).
Again, as the peak of the observed VLM
binary separation distribution falls within this range,
it is possible that the Bate et al.\ simulations underpredict the number
of VLM binary systems.  Because the simulations are computationally
expensive, only one simulation is undertaken for a given set of
initial conditions, resulting in poor statistics.  In addition, the
simulations are allowed to run for a limited
time ($\sim$0.3~Myr), so long term evolution of unstable multiples is left unresolved.

More recent SPH + N-body simulations have attempted to tackle these issues
by reducing the scale of the simulation. Studies by
{\it Delgado-Donate et~al.}~(2004) and {\it Goodwin et al.}~(2004a,b)
have focused on smaller clouds ($\sim5~M_{\odot}$) and have
performed multiple simulations to improve statistical results.
The {\it Delgado-Donate et al.}~simulations were based on the same format
as the Bate et al.\ work and proceeded in two steps;
first an SPH + N-body simulation of the gas and sink particles
was conducted for $\sim$0.5 Myr, followed by an N-body simulation of the resulting
protostellar cores for a subsequent 10 Myr.
This allows an examination of both early fragmentation and accretion on
the formation and disruption of bound systems, and the
dynamical relaxation of high-order multiples over time.
While brown dwarfs were frequently found in multiple systems containing
more massive stellar components, particularly at early times ($\sim$1~Myr),
none of the simulations produced purely
VLM binaries, again indicating
a disagreement between theory (or at least the modelling of the theory)
and observations.  A strong trend of binary fraction
with primary mass is found, although this trend is perhaps too strong
(underestimating VLM multiplicity and overestimating stellar multiplicity).
The SPH simulations of {\it Goodwin et al.}~(2004a,b),
which tested variations of the cloud's initial turbulent energy spectrum, also
failed to produce any VLM binaries within 0.3~Myr.  In retrospect, both
sets of simulations may be hindered by their use of 5~AU sink particles and
softened Newtonian potentials, and both groups have
intentions to address these limitations (M.\ Bate, private communication).

Pure N-body simulations have focused on the dynamical evolution of
small-N clusters of protostars, and (because they are less computationally
intensive) have generally produced
more robust statistical predictions for VLM multiples than SPH simulations.
{\it Sterzik and Durisen} (2003) simulated the dynamical interactions of
closely-separated, small-N clusters and were able to
broadly reproduce the empirical trends, including an increasing binary fraction and
median separation with increasing primary mass (cf. Fig.~\ref{fig_avsmass}),
a brown dwarf binary fraction of $\sim$10\%, and a median VLM binary separation of 3~AU.
{\it Umbreit et~al.}~(2005) studied the decay of widely-separated accreting triple
systems (incorporating momentum transfer with N-body dynamics)
and found that VLM systems hundreds of AU apart were
efficiently hardened to a distribution of that
peaks at 3~AU, with a long tail to wider separations.  These simulations
predict few very tight brown dwarf binary systems, although this may be
because dissipative forces were not included.
One drawback to both of these studies is that they do not take into
account interactions with the larger star-forming environment,
which appear to be important in the SPH simulations of, e.g., {\it Bate et al.}
There are plans to study these effects in detail
(S.\ Umbreit, private communication).

In short, dynamical simulations appear to reproduce many of the observed properties
of VLM binaries, both in terms of quantitative results (binary fraction and separation
distribution) and overall trends (mass dependence on binary fraction and mean separations).
SPH + N-body simulations, on the other hand, generally underpredict
the number of VLM binaries, and the lack of statistics makes the assessment
of other multiple properties difficult to verify.
The shortcomings of SPH simulations are likely related
to the use of large sink particles.  Decreasing the size of these
sink particles, and the imposed smooth potential for close interactions,
should be a priority.

\bigskip
\noindent
\textbf{3.2 Other Formation Mechanisms}
\bigskip

For completeness, we briefly touch upon two other modes of star formation
that may be relevant to the creation of VLM multiples.
Disk fragmentation can occur when
gravitational instabilities in massive circumstellar disks form,
either through dynamical interactions with
a passing bare star or another disk, or spontaneously through
tidal or spiral instabilities.
Most disk fragmentation simulations use SPH
codes to test the outcomes of different encounter
geometries, with results depending largely
on the alignment of the angular
momentum axes of the interacting pair.  The simulations of
{\it Lin et al.}~(1998); {\it Watkins et al.}~(1998a,b); and {\it Boss} (2000) have
all successfully produced substellar mass objects through this
process (the simulations of {\it Bate et al.}~have also produced protostellar cores
through disk interactions).
However, the disk mass necessary to produce such objects is
nearly $0.1~M_{\odot}$, and is hence unlikely around a VLM
primary. Therefore, while
the disk fragmentation scenario appears quite capable of producing single brown
dwarfs from disks around massive stars,
it does little to explain the production of VLM
binary systems.

Another VLM formation
mechanism recently explored by {\it Whitworth and Zinnecker} (2004)
is photo-evaporation.  This process occurs when a substantial
prestellar core (a few 0.1~M$_{\sun}$) is compressed and stripped
by the ionizing radiation front of a nearby massive O or B star.
{\it Whitworth and Zinnecker} (2004) do not discuss binary formation explicitly,
but is possible in principle if the initial core was fragmented.
This scenario also requires the presence of massive young stars,
making it appropriate for high-mass star formation
environments such as Orion, but not for
low-mass environments such as Taurus or Cha~I.  Therefore, photoevaporation
cannot be a universal mechanism for VLM multiple formation.

\section{\textbf{FUTURE OBSERVATIONAL DIRECTIONS}}

Despite the the large
assemblage of VLM binaries now in place (Table~1),
it should be clear that the search for VLM binaries should
continue, particularly by broadening the multiplicity parameter space sampled.
As such, search efforts should focus on low mass ratio systems ($q \lesssim 0.5$),
particularly in the field; very tight
systems ($\Delta \lesssim 3$~AU); and very wide systems ($\Delta \gtrsim 150$~AU)
with moderate to low mass ratios ($q \lesssim 0.8$).

High resolution imaging will remain an important tool in the
discovery and characterization of VLM binaries,
particularly with the implementation of laser guide star (LGS)
AO systems on 5-10~m class telescopes (e.g., Palomar, Keck, VLT).
LGS AO greatly increases the number of VLM systems that are
accessible from the ground.
Ground-based AO
enables the examination of larger samples in the field and in
nearby moving groups and young star forming regions; and the ability
to astrometrically monitor systems on decadal time periods, long
after $HST$ is decommissioned.
Future studies combining AO imaging with spectroscopy
will permit refined characterization of VLM binary components;
note that most of the systems listed in Table~1 lack
resolved spectroscopy.
AO plus coronagraphy, the latter used successfully
to identify several
VLM companions to nearby, more massive stars
(e.g., {\it Oppenheimer et al.}, 2001, {\it Lowrance et al.}, 2005)
will facilitate the detection of low mass ratio systems
around VLM primaries, probing well into the so-called ``planetary mass''
regime.

For the tightest binaries, high resolution spectroscopy remains
an important tool for search and characterization.
Efforts thus far
have been largely conducted at optical wavelengths.  While suitable for young
brown dwarfs with M spectral types, optical spectroscopy
becomes increasingly limited for L dwarfs, T dwarfs and cooler
objects which are extremely faint at these wavelengths.
Hence, future studies should focus their efforts using high-throughput,
high-resolution infrared spectrographs (e.g., {\it Simon and Prato}, 2004).
Short- and long-term spectroscopic monitoring campaigns
of VLM samples should
be pursued to identify sufficiently complete samples and to
determine systemic properties.
Of the few RV variable VLM binary candidates
identified to date
({\it Guenther and Wuchterl}, 2003; {\it Kenyon et al.}, 2005; {\it Joergens}, 2006),
most have only 2--4 epochs of observation, and
parameters such as separation, mass ratio, etc., remain largely unknown.
Combining astrometric monitoring with spectroscopic
monitoring for closely-separated resolved systems
(e.g., Gliese~569Bab; {\it Zapatero Osorio et al.}, 2004)
will permit precise orbital solutions, leading to
component mass and semimajor axis measurements, and
enabling the examination of other multiplicity properties such as eccentricity distributions and
spin/orbit angular momentum alignment.

Tight binaries can also be probed by searches for eclipsing systems.
For substellar objects, this is a particularly powerful technique, as the
near-constancy of evolved (i.e., field) brown dwarf radii
over a broad range of masses
({\it Burrows and Liebert}, 1993) implies that eclipse depths for edge-on
geometries depend only on the relative fluxes of the components, while
grazing transits can span a larger range of inclinations for a given separation.
To date, only one eclipsing substellar system has been identified in
the $\sim$1~Myr ONC,
2MASS~J0535218-054608 ({\it Stassun et al.}, 2006).
To the best of our knowledge no large surveys for eclipsing
field VLM binaries have been undertaken.  While eclipsing systems will likely be rare,
the success and scientific yield of transiting extrasolar planet
searches (e.g., {\it Charbonnaeu et al.}, 2000) should inspire dedicated programs
in this direction.

Interferometric observations can also probe tighter binaries
than direct imaging, encouraging studies in this
direction.  Current facilities (e.g., Palomar, Keck, VLT)
are limited in sensitivity, however; only the closest
mid-type M dwarfs have been observed thus far ({\it Lane et al.}, 2001a;
{\it Segransan et al.}, 2003).
Increasing the throughput of these systems, or making use
of future space-based facilities (e.g., SIM, TPF-I), may eventually
make interferometry a viable observational method in the VLM reg\'ime.

For widely-separated VLM companions, the most extensive limits
to date arise from the shallow, wide-field surveys from which most
of these objects were identified (e.g., 2MASS, DENIS and SDSS).
Only a few dedicated wide-field
programs are now underway ({\it Bill\`eres et al.}, 2005; {\it Allen et al.}~in preparation).
Deep, but not necessarily high resolution imaging surveys
around large samples of VLM primaries would provide better constraints
on the frequency and properties of such systems.
Such surveys will benefit from proper motion analysis and
component spectroscopy, allowing bona-fide systems to be extracted from
the vast number of unrelated projected doubles.
Searches for wide companions to young nearby
stars have identified a few very low mass objects
(e.g., {\it Chauvin et al.}, 2005b; {\it Neuh\"auser et al.}, 2005),
and the case of 2MASS 1207-3934AB proves that widely separated
low mass companions can exist around VLM primaries.
Future searches for equivalent systems, particularly in the field,
will test the veracity of the apparent wide-separation desert.

Finally, careful selection of binary search samples should
be of high priority.  Current imaging and spectroscopic field
samples are largely based on compilations from magnitude-limited
surveys, and are therefore inherently biased.  The examination
of {\em volume-limited} VLM samples (e.g., {\it Cruz et al.}, 2002)
is necessary to eliminate these biases.  Similarly, many cluster
binary surveys fail to concurrently verify cluster membership, leading to
contamination issues (e.g., CFHT-Pl-18; {\it Mart\'in et al.}, 1998, 2000a).
Studies have begun to address this (e.g., {\it Luhman}, 2004), but more
work is needed.  Finally, given the suggestion of age and/or environmental
effects in binary
properties, comparison of large, complete samples for several clusters
of different ages will probe the origins of multiplicity properties
and over what timescales VLM binaries evolve.

\textbf{ Acknowledgments.} This review would not be possible without
helpful input from several researchers that have contributed to the study
of VLM stars and brown dwarf binaries.
We acknowledge specific discussions with
G.\ Basri, I.\ Bonnell, A.\ Boss, G.\ Duch{\^{e}}ne, T.\ Forveille,  R.\ Jeffries,
V.\ Joergens, C.\ Lada, M.\ Liu, K.\ Luhman, E.\ Mart\'in,
E.\ Mamajek, S.\ Metchev, K.\ Noll, P.\ Padoan, D.\ Pinfield, B.\ Reipurth,
K.\ Stassun, M.\ Sterzik, S.\ Umbreit,
R.\ White and A.\ Whitworth.
We also thank our anonymous referee for her/his insightful criticisms.
Support for this work has been provided by NASA through funding
for $Hubble~Space~Telescope$ program GO-10559.

\bigskip

\centerline\textbf{ REFERENCES}
\bigskip
\parskip=0pt
{\small
\baselineskip=11pt

\refs Abt H.\ A. (1978) In {\it Protostars and Planets} (T.\ Gehrels, ed.), pp. 323-355, Univ.\ of Arizona, Tucson.
\refs Abt H.\ A. and Levy S.\ G. (1976) \apjs, {\it 30}, 273-306.
\refs Allen P.~R., Koerner D.~W., Reid I.~N., and Trilling D.~E. (2005) \apj, {\it 625}, 385-397
\refs Ardila D., Mart\'in E., and Basri G. (2000) \aj, {\it 120}, 479-487.
\refs Basri G. (2000) \araa, {\it 38}, 485-519.
\refs Basri G. and Mart\'in E. (1999) \aj, {\it 118}, 2460-2465.
\refs  Bate M.~R. and Bonnell I.~A. (2005) \mnras, {\it 356}, 1201-1221.
\refs  Bate M.~R., Bonnell I.~A., and Bromm V. (2002) \mnras, {\it 332}, L65-L68.
\refs  Bate M.~R., Bonnell I.~A., and Bromm V. (2003) \mnras, {\it 339}, 577-599.
\refs Bill\`eres M., Delfosse X., Beuzit J.-L., Forveille T., Marchal L., et al. (2005) \aap, {\it 440}, L55-L58.
\refs  Boss A.~P. (2000) \apjl, {\it 536}, L101-L104.
\refs  Boss A.~P. (2001) \apjl, {\it 551}, L167-L170.
\refs  Boss A.~P. (2002) \apj, {\it 568}, 743-753.
\refs  Boss A.~P. (2004) \mnras, {\it 350}, L57-L60.
\refs Bouy H., Brandner W., Mart\'in E.~L., Delfosse X., Allard F., et al. (2003), \aj, {\it 126}, 1526-1554.
\refs Bouy H., Brandner W., Mart\'in E.~L., Delfosse X., Allard F., et al. (2004a) \aap, {\it 423}, 341-352.
\refs Bouy H., Brandner W., Mart{\'{\i}}n E.\ L., Delfosse X., Allard F., et al. (2004b) \aap, {\it 424}, 213-226.
\refs Bouy H., Mart{\'{\i}}n E.\ L., Brandner W., and Bouvier J. (2005) \aj, {\it 129}, 511-517.
\refs Bouy H., Moraux E., Bouvier J., Brandner W., Mart\'in E.\ L., et al. (2006) \apj, in press.
\refs  Boyd D.~F.~A. and Whitworth A.~P.\ (2005) \aap, {\it 430}, 1059-1066.
\refs Brandner W., Mart\'in E.~L., Bouy H., K\"ohler R., Delfosse X., et al. (2004) \aap, {\it 428}, 205-208.
\refs Brice\~{n}o C., Hartmann L., Stauffer J., and Mart\'in E. (1998) \aj, {\it 115}, 2074-2091.
\refs Burgasser A.\ J. and McElwain M.\ W. (2006) \aj, in press.
\refs Burgasser A.\ J., Kirkpatrick J.\ D., Brown M.\ E., Reid I.\ N., Burrows A., et al. (2002) \apj, {\it 564}, 421-451.
\refs Burgasser A.\ J., Kirkpatrick J.\ D., Brown M.\ E., Reid I.\ N., Gizis J. E., et al. (1999) \apjl, {\it 522}, L65-L68.
\refs Burgasser A.\ J., Kirkpatrick J.\ D., and Lowrance P.~J. (2005a) \aj, {\it 129}, 2849-2855.
\refs Burgasser A.\ J., Kirkpatrick J.\ D., Reid I.\ N., Brown M.\ E., Miskey C.\ L., et al. (2003) \apj, {\it 586}, 512-526.
\refs Burgasser A.\ J., Reid I.\ N., Leggett S.\ K., Kirkpatrick J.\ D., Liebert, J., et al. (2005b) \apj, {\it 634}, L177-L180.
\refs Burrows A. and Liebert J. (1993) {\it Rev.~Mod.~Phys., 65}, 301-336.
\refs Burrows A., Hubbard W.~B., Lunine J.~I., and Liebert J. (2001) {\it Rev.~Mod.~Phys., 73}, 719-765.
\refs Casey B.~W., Mathieu R.~D., Vaz L.~P.~R., Andersen J., and Suntzeff N.~B. (1998) \aj, {\it 115}, 1617-1633.
\refs Chabrier G., Baraffe I., Allard F., and Hauschildt P. (2000) \apj, {\it 542}, 464-472.
\refs Chappelle R.~J., Pinfield D.~J., Steele I.~A., Dobbie P.~D., and Magazz\'u A. (2005) \mnras, {\it 361}, 1323-1336.
\refs Charbonneau D., Brown T.\ M., Latham D.\ W., and Mayor M. (2000) \apj, {\it 529}, L45-L48.
\refs Chauvin G., Lagrange A.-M., Dumas C., Zuckerman B., Mouillet D., et al. (2004) \aap, 425, L29-L32.
\refs Chauvin G., Lagrange A.-M., Dumas C., Zuckerman B., Mouillet D., et al. (2005a) \aap, 438, L25-L28.
\refs Chauvin G., Lagrange A.-M., Zuckerman B., Dumas C., Mouillet D., et al. (2005b) \aap, 438, L29-L32.
\refs Close L.~M., Richer H.~B., and Crabtree D.~R. (1990) \aj, {\it 100}, 1968-1980.
\refs Close L.~M., Siegler N., Freed M., and Biller B. (2003) \apj, {\it 587}, 407-422.
\refs Close L.\ M., Siegler N., Potter D., Brandner W., and Liebert J. 2002, \apj, {\it 567}, L53-L57.
\refs Cruz K.~L., Reid I.~N., Liebert J., Kirkpatrick J.~D., and Lowrance P.~J. (2003) \aj, {\it 126}, 2421-2448.
\refs Dahn C.\ C., Liebert J., and Harrington R.~S. (1986) \aj, {\it 91}, 621-625.
\refs Dahn C.\ C., Harris H.~C., Vrba F.~J., Guetter H.~H., Canzian B., et al. (2002) \aj, {\it 124}, 1170-1189.
\refs de Zeeuw P.~T., Hoogerwerf R., de Bruijne J.~H.~J., Brown A.~G.~A., and Blaauw A. (1999) \aj, {\it 117}, 354-399.
\refs Delfosse X., Tinney C. G., Forveille T., Epchtein N., Bertin, E., et al. (1997) \aap, {\it 327}, L25-L28.
\refs Delfosse X., Beuzit J.-L., Marchal L., Bonfils X.\ C., Perrier, C., et al. (2004).
In {\it Spectroscopically and Spatially Resolving the Components of the Close Binary Stars} (R.\ W. Hidlitch et al.), pp.\ 166-174. ASP, San Francisco.
\refs Delgado-Donate D.~J., Clarke C.~J., Bate M.~R., and Hodgkin S.~T. (2004) \mnras, {\it 351}, 617-629.
\refs Duch\^ene G., Bouvier J., Bontemp S., Andr\'e P, and Motte F. (2004) \aap, {\it 427}, 651-665.
\refs Duquennoy A. and Mayor M. (1991) \aap, {\it 248}, 485-524 (DM91).
\refs Elmegreen B.\ G. (1997) \apj, {\it 486}, 944-954.
\refs Elmegreen B.\ G. (1999) \apj, {\it 527}, 266-284.
\refs Elmegreen B.\ G. (2000) \apj, {\it 530}, 277-281.
\refs Fischer D.~A. and Marcy G.~W. (1992) \apj, {\it 396}, 178-194 (FM92).
\refs Forveille T., Beuzit J.-L., Delorme P., S\'egransan D., Delfosse X., et al. (2005), \aap, {\it 435}, L5-L9.
\refs Freed M., Close L., and Siegler N. (2003) \apj, {\it 584}, 453-458.
\refs Geballe T.\ R., Knapp G. R., Leggett S. K., Fan, X., Golimowski D. A., et al. (2002) \apj, {\it 564}, 466-481.
\refs Ghez A.~M., Neugebauer G., and Matthews K. (1993) \aj, {\it 106}, 2005-2023.
\refs Gizis J.\ E. (2002), \apj, {\it 575}, 484-492.
\refs Gizis J.~E. and Reid I.~N. (2000) {\it PASP}, {\it 112}, 610-613.
\refs Gizis J.~E., Monet D.~G., Reid I.~N., Kirkpatrick J.~D., and Burgasser A.\ J. (2000a), \mnras, {\it 311}, 385-388.
\refs Gizis J.~E., Monet D.~G., Reid I.~N., Kirkpatrick J.~D., Liebert J., et al. (2000b), \aj, {\it 120}, 1085-1099.
\refs Gizis J.~E., Kirkpatrick J.~D., Burgasser A.~J., Reid I.~N., Monet D.~G., et al. (2001) \apj, {\it 551}, L163-L166.
\refs Gizis J.~E. Reid, I.~N. Knapp, G.~R. Liebert, J. Kirkpatrick, J.~D., et al. (2003) \aj, {\it 125}, 3302-3310.
\refs Goldberg D., Mazeh T., and Latham D.~W. (2003) \apj, {\it 591}, 397-405.
\refs Golimowski D.\ A., Henry T.\ J., Krist J.\ E., Dieterich S., Ford H.\ C., et al. (2004), \aj, 128, 1733-1747.
\refs Goodwin S.~P., Whitworth A.~P., and Ward-Thompson D. (2004a) \aap, {\it 414}, 633-650.
\refs Goodwin S.~P., Whitworth A.~P., and Ward-Thompson D. (2004b) \aap, {\it 423}, 169-182.
\refs Goto M., Kobayashi N., Terada H., Gaessler W., Kanzawa T., et al. (2002) \apj, {\it 567}, L59-L62
\refs Guenther E.~W., Paulson D.~B., Cochran W.~D., Patience J., Hatzes A.~P., et al. (2005) \aap, {\it 442}, 1031-1039.
\refs Guenther E.~W., and Wuchterl G. (2003) \aap, {\it 401}, 677-683
\refs Haisch K.\ E.\ Jr., Barsony M., Greene T.\ P., and Ressler M.\ E. (2002) \aj, {\it 124}, 2841-2852.
\refs Haisch K.\ E.\ Jr., Greene T.\ P., Barsony M., and Stahler S.\ W. (2004) \aj, {\it 127}, 1747-1754.
\refs Halbwachs J.\ L., Mayor M., Udry S., and Arenou F. (2003) \aap, {\it 397}, 159-175.
\refs Henriksen R.~N. (1986) \apj, {\it 310}, 189-206.
\refs Henriksen R.~N. (1991) \apj, {\it 377}, 500-509.
\refs Henry T.\ J. and McCarthy D.\ W.\ Jr. (1990) \apj, {\it 350}, 334-347.
\refs Hillenbrand L.\ A. (1997) \aj, {\it 113}, 1733-1768.
\refs Hinz J.~L., McCarthy D.~W., Simons D.~A., Henry T.~J., Kirkpatrick J.~D., et al. (2002) \aj, {\it 123}, 2027-2032.
\refs Joergens V. (2006) In {\it PPV Poster Proceeedings} \\
http://www.lpi.usra.edu/meetings/ppv2005/pdf/8034.pdf
\refs Joergens V. and Guenther E.\ W. (2001) \aap, {\it 379}, L9-L12.
\refs Kendall T.\ R., Delfosse X., Mart{\'{\i}}n E.\ L., and Forveille T. (2004) \aap, {\it 416}, L17-L20.
\refs Kenworthy M., Hofmann K.-H., Close L., Hinz, P., Mamajek E., et~al. (2001) \apj, {\it 554}, L67-L70.
\refs Kenyon S.\ J., Dobrzycka D., and Hartmann L. (1994) \aj, {\it 108}, 1872-1880.
\refs Kenyon M.~J., Jeffries R.~D., Naylor T., Oliveira J.~M. and Maxted P.~F.~L. (2005) \mnras, {\it 356}, 89-106.
\refs Kirkpatrick J.\ D. (2005) \araa, {\it 43}, 195-245.
\refs Kirkpatrick J.\ D. Reid I.\ N., Liebert J., Cutri R.\ M., Nelson B., et al. (1999) \apj, {\it 519}, 802-833.
\refs Kirkpatrick J.~D., Reid I.~N., Liebert J., Gizis J.~E., Burgasser A.~J., et al. (2000) \aj, {\it 120}, 447-472.
\refs Koerner D.~W., Kirkpatrick J.~D., McElwain M.~W., and Bonaventura N.~R. (1999) \apj, {\it 526}, L25-L28.
\refs Kouwenhoven M.~B.~N., Brown A.~G.~A., Zinnecker H., Kaper L., and Portegies Zwart S.~F. (2005) \aap, {\it 430}, 137-154.
\refs Kraus A.~L., White R.~J., and Hillenbrand L.~A. (2005) \apj, {\it 633}, 452-459.
\refs Kroupa P. (1995a) \mnras, {\it 277}, 1491-1506
\refs Kroupa P. (1995b) \mnras, {\it 277}, 1507-1521
\refs Kroupa P. (1995c) \mnras, {\it 277}, 1522-1540
\refs Kroupa P. and Bouvier J. (2003) \mnras, {\it 346}, 343-353.
\refs Kroupa P., Petr M.\ G., and McCaughrean M.\ J. (1999) {\it New Astron.}, {\it 4}, 495-520.
\refs Lada C.\ J. and Lada E.\ A. (2003) \araa, {\it 41}, 57-115.
\refs Lane B.~F., Boden A.~F., and Kulkarni S.~R. (2001a) \apj, {\it 551}, L81-L83.
\refs Lane B.~F., Zapatero Osorio M.~R., Britton M.~C., Mart{\'{i}}n E.~L., and Kulkarni S.~R. (2001b) \apj, {\it 560}, 390-399.
\refs Larson R.~B. (1992) \mnras, {\it 256}, 641-646.
\refs Law N.\ M., Hodgkin S.\ T., and Mackay C.\ D. 2006, \mnras, in press.
\refs Leggett S.\ K., Golimowski D.\ A., Fan X., Geballe T.\ R., Knapp G.\ R., et al. (2002) \apj, {\it 564}, 452-465.
\refs Leinert Ch., Allard F., Richichi A., and Hauschildt P.\ H. (2000) \aap, {\it 353}, 691-706.
\refs Leinert Ch., Haas M., Mundt R., Richichi A., and Zinnecker H. (1991) \aap, {\it 250}, 407-419.
\refs Leinert Ch., Jahreiss H., Woitas J., Zucker S., Mazeh T., et al. (2001) \aap, {\it 367}, 183-188.
\refs L\'epine S. and Shara M.\ M. (2005) \aj, {\it 129}, 1483-1522.
\refs Lin D.~N.~C., Laughlin G., Bodemheimer P., and Rozyczka M. (1998) {\it Science}, {\it 281}, 2025-2027.
\refs Liu M.\ C. and Leggett S.\ K. (2005) \apj, {\it 634}, 616-624.
\refs Lowrance P.\ J., Becklin E. E., Schneider G., Kirkpatrick J.\ D., Weinberger A.\ J., et al. (2005) \aj, {\it 130}, 1845-1861.
\refs Lucas P.~W., Roche P.~F., and Tamura M. (2005) \mnras, {\it 361}, 211-232.
\refs Luhman K.~L. (2004) \apj, {\it 614}, 398-403.
\refs Luhman K.\ L., McLeod K.\ K., and Goldenson N. (2005) \apj, {\it 623}, 1141-1156.
\refs Mamajek E.\ E. (2005) \apj, {\it 634}, 1385-1394.
\refs Marcy G.\ W. and Butler R.\ P. (2000) \pasp, {\it 112}, 137-140.
\refs Mart\'{\i}n E.~L., Brandner W., and Basri G. (1999) {\it Science}, {\it 283}, 1718-1720.
\refs Mart{\'{\i}}n E.\ L., Brandner W., Bouvier J., Luhman K.\ L., Stauffer J., et al. (2000a) \apj, {\it 543}, 299-312.
\refs Mart\'{\i}n E.~L., Barrado y Navascu\'{e}s D., Baraffe I., Bouy H., and Dahm S. (2003) \apj, {\it 594}, 525-532.
\refs Mart{\'{\i}}n E.\ L., Koresko C.\ D., Kulkarni S.\ R., Lane B.\ F., and Wizinowich P.\ L. (2000b) \apjl, {\it 529}, L37-L40.
\refs Mart\'in E.\ L., Rebolo R., and Zapatero Osorio M.\ R. (1996) \apj, {\it 469}, 706
\refs Mart\'{\i}n E.~L., Basri G., Brandner W., Bouvier J., Zapatero Osorio M. R., et al. (1998) \apj, {\it 509}, L113-L116
\refs Maxted P.~F.~L. and Jeffries R.~D. (2005) \mnras, {\it 362}, L45-L49.
\refs Mayor M., Duquennoy A., Halbwachs J.~L., and Mermilliod J.~C. (1992) {\it In Complementary Approaches to Double and Multiple Star Research, IAU Colloquium 135} (H.\ MacAlister and W.~I.\ Hartkopf, eds.), pp.\ 73-81. ASP, San Francisco
\refs Mazeh T. and Goldberg D. (1992) \apj, {\it 394}, 592-598.
\refs McCaughrean M., Close L.\ M., Scholz R.-D., Lenzen R., Biller B., et al. (2004) \aap, {\it 413}, 1029-1036.
\refs McGovern M.\ R. (2005) Ph.D.\ Thesis, UC Los Angeles.
\refs McLean I.\ S., McGovern M.\ R., Burgasser A.\ J., Kirkpatrick J.\ D., Prato L., et al. (2003) \apj, {\it 596}, 561-586.
\refs Metchev S.\ A. (2005) Ph.D.\ Thesis, California Institute of Technology.
\refs Neuh\"auser R., Brandner W., Alves J., Joergens V., and Comer\'on F. (2002) \aap, {\it 384}, 999-1011.
\refs Neuh\"{a}user R., Guenther E.~W., Alves J., Hu{\'e}lamo N., Ott T., et al. (2003) {\it AN}, {\it 324}, 535-542.
\refs Neuh\"auser R., Guenther E.~W., Wuchterl G., Mugrauer M., Bedalov A., et al. (2005) \aap, {\it 435}, L13-L16.
\refs Oppenheimer B.~R., Golimowski D.~A., Kulkarni S.~R., Matthews K., Nakajima T., et al. (2001) \aj, {\it 121}, 2189-2211.
\refs Oppenheimer B.~R., Kulkarni S.~R. and Stauffer J.~R. (2000) In {\it Protostars and Planets IV} (V.\ Mannings et al., eds.) p. 1313. Univ.\ of Arizona, Tucson,
\refs Padoan P., Kritsuk A., Norman M.~L., and Nordlund A. (2005) {\it Mem.\ S.A.It.,76}, 187-192.
\refs Padoan P. and Nordlund A. (2004) \apj, {\it 617}, 559-564.
\refs Patience J., and Duch\^ene G. (2001) In {\it IAU Symp.\ 200, The Formation of Binary Stars} (H.\ Zinnecker and R.\ D.\ Mathieu, eds.) p. 181. ASP, San Francisco.
\refs Percival S.~M., Salaris M., and Groenewegen M.\ A.\ T. (2005) \aap, {\it 429}, 887-894.
\refs Perryman M. A. C., Lindegren L., Kovalevsky J., Hoeg E., Bastian U., et al. (1997) \aap, {\it 323}, L49-L52.
\refs Pinfield D.~J., Dobbie P.~D., Jameson R.~F., Steele I.~A., Jones H.~R.~A., et al. (2003) \mnras, {\it 342}, 1241-1259.
\refs Potter D., Mart\'in E.\ L., Cushing M.\ C., Baudoz P., Brandner W., et al. (2002) \apj, {\it 567}, L133-L136.
\refs Reid I.~N. and Gizis J.~E. (1997a) \aj, {\it 113}, 2246-2269.
\refs Reid I.~N. and Gizis J.~E. (1997b) \aj, {\it 114}, 1992-1998.
\refs Reid I.~N., Gizis J.~E., and Hawley S.~L. (2002a) \aj, {\it 124}, 2721-2738.
\refs Reid I.~N., Gizis J.~E., Kirkpatrick J.~D., Koerner D.~W. (2001) \aj, {\it 121}, 489-502.
\refs Reid I.~N., Kirkpatrick J.~D., Liebert J., Gizis J.~E., Dahn C.~C., et al. (2002b) \aj, {\it 124}, 519-540.
\refs Reid I.~N., Lewitus E., Burgasser A.\ J., and Cruz K.\ L. (2006) \apj, in press
\refs Reid I.~N., Kirkpatrick J.\ D., Liebert J., Burrows A., Gizis, J.\ E., et al. (1999) \apj, {\it 521}, 613-629.
\refs Reid I.~N., Cruz K.\ L., Laurie S.\ P., Liebert J., Dahn C.\ C., et al. (2003) \aj, {\it 125}, 354-358.
\refs Reipurth B. and Clarke C. (2001) \aj, {\it 122}, 432-439.
\refs Reipurth B., Rodr\'iguez L.\ F., Anglada G., and Bally J. (2002) \aj, {\it 124}, 1045-1053.
\refs Reipurth B., Rodr\'iguez L.\ F., Anglada G., and Bally J. (2004) \aj, {\it 127}, 1736-1746.
\refs Scally A., Clarke C., and McCaughrean M.\ J. (1999) \mnras, {\it 306}, 253-256.
\refs Schmidt M. (1968) \apj, {\it 151}, 393-409.
\refs S\'egransan D., Kervella P., Forveille T., and Queloz D. (2003) \aap, {\it 397}, L5-L8.
\refs Seifahrt A., Guenther E., and Neuh\"auser R. (2005) \aap, {\it 440}, 967-972.
\refs Shatsky N. and Tokovinin A. (2002) \aap, {\it 382}, 92-103.
\refs Shu F.\ H., Adams F.\ C., and Lizano S. (1987) \araa, {\it 25}, 23-81.
\refs Siegler N., Close L.~M., Mamajek E.~E., and Freed M. (2003) \apj, {\it 598}, 1265-1276.
\refs Siegler N., Close L.\ M., Cruz K.\ L., Mart{\'{\i}}n E.\ L., and Reid I.\ N. (2005) \apj, {\it 621}, 1023-1032.
\refs Simon M. and Prato L. (2004) \apj, {\it 613}, L69-L71.
\refs Sivia D. (1996) {\it Data Analysis} Oxford: Claredon Press
\refs Stassun K., et~al. (2006) {\it Nature}, in press 
\refs Stauffer J.\ R., Schultz G., and Kirkpatrick J.\ D. (1998) \apj, {\it 499}, L199-L203.
\refs Stephens D.\ C., Marley M.\ S., Noll K.\ S., and Chanover N. (2001) \apj, {\it 556}, L97-L101.
\refs Sterzik M.~F. and Durisen R.~H. (2003) \aap, {\it 400}, 1031-1042.
\refs Tinney C.\ G. (1996) \mnras, {\it 281}, 644-658.
\refs Tinney C.\ G., Burgasser A.\ J., and Kirkpatrick J.\ D. (2003) \aj, {\it 126}, 975-992.
\refs Tokovinin A.\ A. (1997) {\it Astron.\ Astrophys.\ Supp., 124}, 71-76.
\refs Tokovinin A.\ A. (2004) \rmxaa, {\it 21}, 7-14.
\refs Udry S., Mayor M., and Queloz D. (2003) In {\it ASP Conf.~Ser.~294: Scientific Frontiers in Research on Extrasolar Planets} (D.\ Deming and S.\ Seager, eds.) pp.\ 17-26. ASP: San Francisco.
\refs Umbreit S., Burkert A., Henning T., Mikkola S., \& Spurzem R. (2005) \apj, {\it 623}, 940-951.
\refs van Altena W.\ F., Lee J.\ T., and Hoffleit E.\ D. (1995) {\it The General Catalog of Trignometric Stellar Parallaxes, 4$^{th}$ Edition} Yale Univ.\ Obs.: New Haven.
\refs Vrba F.\ J., Henden A.\ A., Luginbuhl C.\ B., Guetter H.\ H., Munn J.\ A., et al. (2004), \aj, {\it 127}, 2948-2968.
\refs Watkins S.~J., Bhattal A.\ S., Boffin H.\ M.\ J., Francis N., and Whitworth A.\ P. (1998a) \mnras, {\it 300}, 1205-1213.
\refs Watkins S.~J., Bhattal A.\ S., Boffin H.\ M.\ J., Francis N., and Whitworth A.\ P. (1998b) \mnras, {\it 300}, 1214-1224.
\refs Weinberg M.\ D., Shapiro S.\ L., and Wasserman I. (1987) \apj, {\it 312}, 367-389.
\refs White R.\ D., Ghez A.\ M., Reid I.\ N., and Schultz G. (1999), \apj, {\it 520}, 811-821.
\refs Whittet D.~C.~B., Prusti T., Franco G.~A.~P., Gerakines P.~A., Kilkenny D., et al. (1997) \aap, {\it 327}, 1194-1205.
\refs Whitworth A.~P. and Zinnecker H. (2004) \aap, {\it 427}, 299-306.
\refs Wilson J.\ C., Kirkpatrick J.\ D., Gizis J.\ E., Skrutskie M.\ F., Monet D.\ G., et al. (2001) \aj, {\it 122}, 1989-2000.
\refs Zapatero Osorio M.~R., Lane B.\ F., Pavlenko Ya., Mart\'in E.\ L., Britton M., et al. (2004) \apj, {\it 615}, 958-971.
\refs Zuckerman B. and Song I. (2004) \araa, {\it 42}, 685-721.

\begin{deluxetable}{lccccccccll}
\tabletypesize{\scriptsize}
\tablecaption{Known Very Low Mass Binaries \label{tbl-1}}
\tablewidth{0pt}
\tablehead{
 & & & &
\multicolumn{2}{c}{Estimated} & &
\colhead{Estimated} & &
\colhead{Association} & \\
\colhead{Source Name} &
\multicolumn{2}{c}{Separation} &
\colhead{Spectral Types} &
\multicolumn{2}{c}{Masses} &
\colhead{$q$} &
\colhead{Period} &
\colhead{Age} &
\colhead{or Note} &
\colhead{Ref.} \\
 &
\colhead{(mas)} &
\colhead{(AU)} &
 &
\colhead{(M$_{\sun}$)} &
\colhead{(M$_{\sun}$)} &
 &
\colhead{(yr)} &
\colhead{(Myr)} &
 &  \\
\colhead{(1)} &
\colhead{(2)} &
\colhead{(3)} &
\colhead{(4)} &
\colhead{(5)} &
\colhead{(6)} &
\colhead{(7)} &
\colhead{(8)} &
\colhead{(9)} &
\colhead{(10)} &
\colhead{(11)} \\
  }
\startdata
Cha H$\alpha$8 & \nodata & \nodata & M6.5 + [M6.5:] & 0.070 & \nodata & \nodata & \nodata & 2 & Cha I; RV & 37 \\
2MASS~J0253202+271333AB & \nodata & \nodata & M8 + [M8:] & 0.092 & 0.092: & 1: & \nodata & \nodata & SB2 & 8; 42 \\
2MASS~J0952219-192431AB & \nodata & \nodata & M7 + [M7:] & 0.098 & 0.098: & 1: & \nodata & \nodata & SB2 & 8; 43 \\
LHS~292AB & \nodata & \nodata & M7 + [M7:] & 0.098 & 0.098: & 1: & \nodata & \nodata & SB2 & 8,28; 76 \\
2MASS~J2113029-100941AB & \nodata & \nodata & M6 + [M6:] & 0.085 & 0.085: & 1: & \nodata & \nodata & SB2 & 28; 42 \\
PPl~15AB & \nodata & 0.03\tablenotemark{a} & M7 + [M8:] & 0.070\tablenotemark{a} & 0.060\tablenotemark{a} & 0.86 & 0.0159\tablenotemark{a} & 120 & Pleiades; SB2 & 1; 47,60 \\
2MASS~J0535218-054608AB  & \nodata & 0.04\tablenotemark{b} & M6.5 + [M6.5] & 0.054\tablenotemark{b} & 0.034\tablenotemark{b}  & 0.63 & 0.0268\tablenotemark{b} &  1  &  Orion; SB2, EB  &  79 \\
2MASS~J15344984-2952274AB & 65 & 0.9 & T5.5 + [T5.5] & 0.035 & 0.035 & 1.00 & 4 & \nodata & & 5; 38 \\
GJ~569BC & 103 & 0.90\tablenotemark{a} & M8.5 + M9.0 & 0.071\tablenotemark{a} & 0.054\tablenotemark{a} & 0.76 & 2.4\tablenotemark{a} & 300 & Ursa Major; triple & 2; 3,33,75,49 \\
GJ~1001BC & 87 & 1.0 & L4.5 + [L4.5] & 0.068 & 0.068 & 1.00 & 4 & \nodata &triple & 25; 35,36,52 \\
LP~349-25AB & 125 & 1.3 & M8 + [M9] & 0.090 & 0.085 & 0.94 & 4 & \nodata & & 31; 42 \\
SDSS~J092615.38+584720.9AB & 70 & 1.4 & T4.5 + [T4.5] & 0.050 & 0.050 & 1.00 & 7 & \nodata & & 69; 71 \\
GJ~417BC & 70 & 1.5 & L4.5 + [L6] & 0.073 & 0.070 & 0.96 & 7 & \nodata & triple & 4; 39,40 \\
2MASS~J0920122+351742AB & 70 & 1.5 & L6.5 + [T:] & 0.068 & 0.068 & 1.00 & 6 & \nodata & & 7; 39,69,78 \\
2MASS~J2252107-173014AB & 140 & 1.9 & L6 + [T2] & 0.070 & 0.060 & 0.86 & 10 & \nodata & & 32; 58,59 \\
2MASS~J1847034+552243AB & 82 & 1.9 & M7 + [M7.5] & 0.098 & 0.094 & 0.96 & 8 & \nodata & & 23; 43 \\
2MASS~J0652307+471034AB &  170  & 2.0 &  L3.5 + [L6.5] & 0.075 &  0.071 & 0.95  &  10 & \nodata  &  &  78; 43  \\
DENIS~PJ035726.9-441730AB & 98 & 2.2 & M9.5 + [L1.5] & 0.085 & 0.080 & 0.91 & 11 & \nodata & & 4,13 \\
HD~130948BC & 134 & 2.4 & L4 + [L4] & 0.070 & 0.060 & 0.86 & 14 & \nodata & triple & 6; 26,40 \\
SDSS~J042348.57-041403.5AB & 164 & 2.5 & L7 + T2 & 0.060 & 0.050 & 0.83 & 16 & \nodata & & 68; 43,70,71 \\
2MASS~J0746425+200032AB & 220 & 2.5\tablenotemark{a} & L0 + L1.5 & 0.085\tablenotemark{a} & 0.066\tablenotemark{a} & 0.78 & 11\tablenotemark{a} & 300 &  &4,7,17; 20,39,41,61,71 \\
$\epsilon$~IndiBC & 732 & 2.6 & T1 + T6 & 0.045 & 0.027 & 0.60 & 22 & 1300 & triple & 16; 40 \\
2MASS~J1430436+291541AB & 88 &  2.6  &  L2 + [L2:] & 0.076 & 0.075 & 0.99  &  15 & \nodata & & 4; 43,86? \\
2MASS~J1728114+394859AB & 131 & 2.7 & L7 + [L8] & 0.069 & 0.066 & 0.96 & 16 & \nodata & & 4,13; 39 \\
LP 213-68AB & 122 & 2.8 & M8 + [L0] & 0.092 & 0.084 & 0.91 & 15 & \nodata & triple & 17,53 \\
LHS~2397aAB & 207 & 3.0 & M8 + [L7.5] & 0.090 & 0.068 & 0.76 & 18 & \nodata & & 10; 36,42,82 \\
LSPM 1735+2634AB & 290 & 3.2 & [M9:] + [M9:] & 0.082 & 0.074 & 0.90 & 14 & \nodata & & 51; 83 \\
LHS~1070BC & 446 & 3.4\tablenotemark{a} & M8.5 + [M9] & 0.070\tablenotemark{a} & 0.068\tablenotemark{a} & 0.97 & 16\tablenotemark{a} & \nodata & quadruple & 18; 74 \\
2MASS~J0856479+223518AB\tablenotemark{c} & 98 & 3.4 & L3: + [L:] & 0.071 & 0.064 & 0.90 & 24 & \nodata &  & 4; 43 \\
2MASS~J1017075+130839AB & 104 &  3.4  &   L2 + [L2] & 0.076  & 0.076 & 1.00  &  23 & \nodata & &   4; 43,86 \\
SDSS~2335583-001304AB & 57 & 3.5 & L1: + [L4:] & 0.079 & 0.074 & 0.94 & 24 & \nodata & & 4; 81 \\
2MASS~J1600054+170832AB & 57 & 3.5 & L1 + [L3] & 0.078 & 0.075 & 0.96 & 23 & \nodata & & 4,13; 39 \\
LP~415-20AB & 119 & 3.6 & M7 + [M9.5] & 0.095 & 0.079 & 0.83 & 22 & 625 & Hyades & 9; 42 \\
2MASS~J12255432-2739466AB & 282 & 3.8 & T6 + [T8] & 0.033 & 0.024 & 0.73 & 43 & \nodata & & 5; 38,77 \\
SDSS~J153417.05+161546.1AB & 106 & 3.8 & T1.5 + [T5.5] & 0.050 & 0.040 & 0.80 & 35 & \nodata & & 15 \\
SDSS~J102109.69-030420.1AB & 160 & 3.9 & T1 + T5 & 0.060 & 0.050 & 0.83 & 33 & \nodata & & 69; 70,72 \\
2MASS~J1426316+155701AB & 152 & 4.0 & M8.5 + [L1] & 0.088 & 0.076 & 0.86 & 27 & \nodata & & 17; 42 \\
2MASS~J2140293+162518AB & 155 & 4.0 & M8.5 + [L2] & 0.092 & 0.078 & 0.85 & 27 & \nodata & & 17; 42 \\
2MASS~J15530228+1532369AB & 340 & 4.4 & T7 + [T7] & 0.040 & 0.030 & 0.75 & 49 & \nodata & & 69; 73 \\
2MASS~J1239272+551537AB & 211 & 4.5 & L5 + [L5] & 0.071 & 0.071 & 1.00 & 35 & \nodata & & 4,13; 39 \\
2MASS~J2206228-204705AB & 168 & 4.5 & M8 + [M8] & 0.092 & 0.091 & 0.99 & 31 & \nodata & & 17; 42 \\
2MASS~J0850359+105716AB & 160 & 4.7 & L6 + [L8] & 0.050 & 0.040 & 0.80 & 39 & \nodata & & 7; 41,52,70 \\
2MASS~J1750129+442404AB & 158 & 4.9 & M7.5 + [L0] & 0.095 & 0.084 & 0.88 & 36 & \nodata & & 9; 42 \\
USco-109AB\tablenotemark{c} & 34 & 4.9 & M6 + [M7.5] & 0.070 & 0.040 & 0.57 & 46 & 5 & Up~Sco & 29; 45,65 \\
2MASS~J2101154+175658AB & 234 & 5.4 & L7 + [L8] & 0.068 & 0.065 & 0.96 & 49 & \nodata & & 4,13; 39 \\
Kelu-1AB & 291 & 5.4 & L2 + [L4] & 0.060 & 0.055 & 0.92 & 52 & \nodata & & 24; 41,52 \\
2MASS~J0429184-312356AB & 531 & 5.8 & M7.5 + [L1] & 0.094 & 0.079 & 0.84 & 48 & \nodata & & 23,78; 43 \\
2MASS~J0147328-495448AB & 190 &  5.8  &  M8 + [M9] & 0.086 & 0.080 & 0.93  &  47  & \nodata   &  & 78 \\
2MASS~J2152260+093757AB & 250 &  6.0  &  L6: + [L6:] &  0.069 & 0.069 & 1.00  &  55  &  \nodata   &  & 78 \\
MHO Tau 8AB & 44 & 6.2 & M6 + [M6.5] & 0.100 & 0.070 & 0.70 & 53 & 2 & Taurus & 55; 56 \\
DENIS~J122815.2-154733AB & 275 & 6.4\tablenotemark{a} & L6 + [L6] & 0.065\tablenotemark{a} & 0.065\tablenotemark{a} & 1.00 & 44\tablenotemark{a} & \nodata & & 11; 41,64,71 \\
DENIS~J100428.3-114648AB &  146 & 6.8 &  L0: + [L2:] & 0.080 & 0.076 & 0.95 &   63 & \nodata & &  4 \\
2MASS~J2147436+143131AB & 322 & 7.0 & M8 + [L0] & 0.084 & 0.078 & 0.93 & 65 & \nodata & & 4,13; 42 \\
DENIS~J185950.9-370632AB & 60 & 7.7 & L0 + [L3] & 0.084 & 0.076 & 0.90 & 76 & 5 & R-CrA & 20; 57 \\
2MASS~J1311391+803222AB & 267 & 7.7 & M8.5 + [M9] & 0.089 & 0.087 & 0.98 & 72 & \nodata & & 17; 42 \\
IPMBD~29AB & 58 & 7.8 & L1 + [L4] & 0.045 & 0.038 & 0.84 & 106 & 120 & Pleiades & 14; 47 \\
2MASS~J1146345+223053AB & 290 & 7.9 & L3 + [L4] & 0.055 & 0.055 & 1.00 & 94 & \nodata & & 7,12; 41,52 \\
CFHT-Pl-12AB & 62 & 8.3 & M8 + [L4] & 0.054 & 0.038 & 0.70 & 111 & 120 & Pleiades & 14; 47,62 \\
2MASS~J1127534+741107AB & 246 & 8.4 & M8 + [M9] & 0.092 & 0.087 & 0.95 & 80 & \nodata & & 17; 42 \\
2MASS~J1449378+235537AB & 134 & 8.5 & L0 + [L3] & 0.084 & 0.075 & 0.89 & 88 & \nodata & & 4,13; 39 \\
LP~475-855AB & 294 & 8.5 & M7.5 + [M9.5] & 0.091 & 0.080 & 0.88 & 85 & 625 & Hyades & 9; 42 \\
DENIS~J020529.0-115925AB &  510 & 9.2 &  L7 + [L7] &  0.070 &  0.070 &    1.00 &   105 &  \nodata  & poss.\ triple &  12; 52,41 \\
USco-66AB & 70 & 10.2 & M6 + [M6] & 0.070 & 0.070 & 1.00 & 120 & 5 & Up~Sco & 29; 45,65 \\
2MASS~J17072343-0558249AB & 950 & 10.4 & M9 + L3 & 0.090 & 0.060 & 0.67 & 125 & \nodata & & 67 \\
GJ~337CD & 530 & 10.9 & L8 + [T:] & 0.055 & 0.055 & 1.00 & 150 & \nodata & quadruple & 30; 50,67 \\
2MASS~J0915341+042204AB & 730 & 11.0 & L7 + [L7] & 0.070 & 0.070 & 1.00 &  138 &  \nodata  & &   78  \\
IPMBD~25AB & 94 & 12.6 & M7 + [L4] & 0.063 & 0.039 & 0.62 & 200 & 120 & Pleiades & 14; 47 \\
DENIS~J144137.3-094559AB & 420 & 14.3 & L1 + [L1] & 0.072 & 0.072 & 1.00 & 200 & \nodata & triple & 4,48; 39,80 \\
2MASS~J2331016-040618AB & 573 & 15.0 & M8 + [L7] & 0.093 & 0.067 & 0.72 & 200 & \nodata & triple & 4,13,17; 42,49 \\
USco-55AB & 122 & 17.7 & M5.5 + [M6] & 0.100 & 0.070 & 0.70 & 250 & 5 & Up~Sco & 29; 45,65 \\
CFHT-Pl-18AB & 330 & 34.6 & M8 + M8 & 0.090 & 0.090 & 1.00 & 680 & \nodata & & 4; 19 \\
DENIS~J220002.0-303832.9AB & 1090 & 38.2 & M8 + L0 & 0.085 & 0.083 & 0.98 & 800 & \nodata & & 66 \\
2MASS~J1207334-393254AB & 776 & 41.1 & M8.5 + L: & 0.024 & 0.004 & 0.17 & 2~250 & 8 & TW~Hyd & 22,63; 34 \\
DENIS~J055146.0-443412.2AB & 2200 & 220.0 & M8.5 + L0 & 0.085 & 0.079 & 0.93 & 11~500 & \nodata & & 27 \\
2MASS~J11011926-7732383AB & 1440 & 241.9 & M7 + M8 & 0.050 & 0.025 & 0.50 & 20~000 & 2 & Cha~I & 21; 44 \\
\enddata
\tablecomments{Uncertain values are indicated by colons.
Additional column information: (1) name of binary; (2) angular
separation in mas;
(3) projected separation ($\Delta$) in AU, or semimajor axis of
orbit as noted; (4) spectral types of binary components; for
sources without resolved spectroscopy, primary spectral type is
for combined light data, secondary spectral type is estimated from
photometric flux ratios (as indicated by brackets); (5) estimated
primary mass in M$_{\sun}$, taken as the average of the reported
mass ranges; masses determined from orbital dynamics are
indicated; (6) estimated secondary mass in M$_{\sun}$, taken as
the average of the reported mass ranges; masses determined from
orbital dynamics are indicated; (7) $q \equiv$ M$_2$/M$_1$, as
reported or calculated from columns [5]-[6]; (8) estimated orbital
period in yr, assuming circular orbit with semimajor axis $a =
1.26\Delta$ (FM92); sources with period measurements from orbital
measurements are indicated; (9) estimated age in Myr of binary if
member of a moving group or association, or companion to a
age-dated star;
(10) additional notes, including cluster association; (11)
references as given below; discovery references are listed first,
followed by references for additional data (spectral types,
distance measurements/estimates, orbital measurements) separated
by a semicolon. } \tablenotetext{a}{Parameters derived or
estimated from orbital motion measurements.}
\tablenotetext{a}{Parameters for 2MASS~J0535218-054608AB based on
both spectroscopic orbit and eclipsing light curve; see {\it Stassun et
al.}~(2006).} \tablenotetext{c}{Candidate binary.}
\tablerefs{ (1) {\it Basri and Mart\'in} (1999); (2) {\it Mart\'in et al.}~(2000b);
(3) {\it Lane et al.}~(2001b); (4) {\it Bouy et al.}~(2003); (5) {\it Burgasser et al.}~(2003);
(6) {\it Potter et al.}~(2002); (7) {\it Reid et al.}~(2001); (8) {\it Reid et al.}~(2002b);
(9) {\it Siegler et al.}~(2003); (10) {\it Freed et al.}~(2003);
(11) {\it Mart\'in et al.}~(1999); (12) {\it Koerner et al.}~(1999);
(13) {\it Gizis et al.}~(2003); (14) {\it Mart\'in et al.}~(2003);
(15) {\it Liu et al.}~(in preparation);
(16) {\it McCaughrean et al.}~(2004);
(17) {\it Close et al.}~(2003); (18) {\it Leinert et al.}~(2001);
(19) {\it Mart\'in et al.}~(2000a); (20) {\it Bouy et al.}~(2004b);
(21) {\it Luhman}~(2004); (22) {\it Chauvin et al.}~(2004);
(23) {\it Siegler et al.}~(2005); (24) {\it Liu and Leggett}~(2005);
(25) {\it Golimowski et al.}~(2004); (26) {\it Goto et al.}~(2002);
(27) {\it Bill\`eres et al.}~(2005); (28) {\it Guenther and Wuchterl}~(2003);
(29) {\it Kraus et al.}~(2005);
(30) {\it Burgasser et al.}~(2005a); (31) {\it Forveille et al.}~(2005);
(32) {\it Reid et al.}~(2006); (33) {\it Kenworthy et al.}~(2001); (34) {\it Mamajek}~(2005);
(35) {\it Leggett et al.}~(2002); (36) {\it van Altena et al.}~(1995);
(37) {\it Joergens}~(2006); (38) {\it Tinney et al.}~(2003);
(39) {\it Kirkpatrick et al.}~(2000);
(40) {\it Perryman}~(1997); (41) {\it Dahn et al.}~(2002); (42) {\it Gizis et al.}~(2000b);
(43) {\it Cruz et al.}~(2003); (44) {\it Whittet et al.}~(1997);
(45) {\it de Zeeuw et al.}~(1999); (46) {\it Kenyon et al.}~(1994);
(47) {\it Percival et al.}~(2005);
(48) {\it Stephens et al.}~(2001); (50) {\it Wilson et al.}~(2001);
(51) {\it Law et al.}~(2006);
(52) {\it Kirkpatrick et al.}~(1999); (53) {\it Gizis et al.}~(2000a);
(54) {\it Delfosse et al.}~(1997); (55) {\it White et al.}~(in preparation);
(56) {\it Brice\~{n}o et al.}~(1998); (57) {\it Casey et al.}~(1998);
(58) {\it Kendall et al.}~(2004); (59) {\it McGovern}~(2005);
(60) {\it Mart\'in et al.}~(1996); (61) {\it Bouy et al.}~(2004a);
(62) {\it Stauffer et al.}~(1998); (63) {\it Chauvin et al.}~(2005a);
(64) {\it Brandner et al.}, 2004; (65) {\it Ardila et al.}~(2000);
(66) {\it Burgasser and McElwain}~(2006); (67) {\it McElwain and Burgasser}~(in preparation);
(68) {\it Burgasser et al.}~(2005b), (69) {\it Burgasser et al.}~(in preparation);
(70) {\it Vrba et al.}~(2004); (71) {\it Geballe et al.}~(2002);
(72) {\it McLean et al.}~(2003); (73) {\it Burgasser et al.}~(2002);
(74) {\it Leinert et al.}~(2000); (75) {\it Zapatero-Osorio et al.}~(2004);
(76) {\it Dahn et al.}~(1986);
(77) {\it Burgasser et al.}~(1999); (78) {\it Reid et al.}~(in preparation);
(79) {\it Stassun et al.}~(2006);
(80) {\it Seifahrt et al.}~(2005);
(81) {\it H.\ Bouy}, private communication (2005);
(82) {\it Tinney}~(1996);
(83) {\it L\'epine and Shara} (2005)
}
\end{deluxetable}

\end{document}